\documentclass[11pt,a4paper]{article}
\usepackage{bm,jcappub,Macro,tabularx}
\usepackage{url}
\usepackage{hyperref}
\usepackage{url}
\usepackage{epstopdf}
\usepackage{enumitem}
\usepackage{multirow}
\usepackage{colortbl}
\usepackage{color}
\usepackage{subfig}
\usepackage{multicol}

\def\la{\langle}
\def\ra{\rangle}
\def\n{\noindent}
\def\be{\begin{equation}}
\def\ee{\end{equation}}
\def\ben{\begin{eqnarray}}
\def\een{\end{eqnarray}}
\def\nn{\nonumber}
\def\oh{\bf \hat\Omega}

\def\bk{{\bf k}}

\def\br{{\bf r}}

\def\rD{{\rm D}}

\def\bk{{\bf k}}

\def\2p{{(2\pi)^2}}

\def\be{\begin{equation}}
\def\ee{\end{equation}}
\def\beq{\begin{equation}}
\def\eeq{\end{equation}}
\def\ben{\begin{eqnarray}}
\def\een{\end{eqnarray}}

\def\oh{{\hat\Omega}}
\def\nn{{\nonumber}}

\newcommand{\beqa}{\begin{eqnarray}}
\newcommand{\eeqa}{\end{eqnarray}}
\newcommand{\rP}{{P}}

\def\ikap0{{\cal J}_{\theta_0}(r)}

\def\one1{\langle \kappa_{(i)}\kappa_{(j)} \rangle}
\def\one{{[\bar \xi^{(ij)}]}}

\def\ba{\begin{eqnarray}}
\def\ea{\end{eqnarray}}

\def\bk{{\bf k}}
\def\bq{{\bf q}}
\def\rL{{\rm L}}
\def\rD{{\rm D}}
\def\rF{{\rm F}}
\def\bell{{\bm\ell}}

\title{The Integrated Bispectrum and Beyond}
\author{Dipak Munshi$^{1}$, Peter Coles$^{2}$}
\affiliation{${}^{1}$Astronomy Centre, School of Mathematical and Physical Sciences,\\ 
University of Sussex, Brighton BN1 9QH, U.K.\\
${}^{2}$School of Physics and Astronomy, Cardiff University, \\
Queen's Buildings, The Parade, Cardiff CF24 3AA, U.K.}
\emailAdd{D.Munshi@sussex.ac.uk, P.Coles@sussex.ac.uk}
\abstract{The position-dependent power spectrum has been recently proposed as a descriptor
of gravitationally induced non-Gaussianity in galaxy clustering, as 
it is sensitive to the ''soft limit" of the bispectrum (i.e. when one of the 
wave number tends to zero). We generalise this concept 
to higher order and clarify their 
relationship to other known statistics such as the skew-spectrum, the kurt-spectra
and their real-space counterparts
the cumulants correlators. Using the {\em Hierarchical Ansatz} (HA) as a toy model for 
the higher order correlation
hierarchy, we show how in the soft limit, polyspectra at a given order
can be identified with lower order polyspectra with the same geometrical dependence
but with {\em renormalised} amplitudes expressed in terms of amplitudes 
of the original polyspectra. 
We extend the concept of position-dependent bispectrum
to bispectrum of the divergence of the velocity field $\Theta$ and mixed multispectra involving
$\delta$ and $\Theta$ in the 3D perturbative regime.
To quantify the effects of transients in
numerical simulations, we also present results for lowest order in
Lagrangian perturbation theory (LPT) or the Zel'dovich approximation (ZA).
Finally, we discuss how to extend the position-dependent
spectrum concept to encompass cross-spectra. And finally study the application
of this concept to two dimensions (2D), for projected galaxy maps, 
convergence $\kappa$ maps from weak-lensing surveys or
maps of CMB secondaries e.g. the frequency cleaned $y$ - parameter maps of thermal
Sunyaev-Zel'dovich (tSZ) effect from CMB surveys.}
\begin{document} 
\maketitle
\section{Introduction}
\label{sec:intro}
Over the last decade advances in astronomical spectroscopy and photometry of large samples of
galaxies have allowed the galaxy distribution to be mapped to unprecedented accuracy and detail. Analysis
of the resulting maps has yielded constraints on the growth rate of structures, expansion
history of the Universe as well as on cosmological parameters. Examples include 
BOSS\footnote{Baryon Oscillator Spectroscopic Survey: \href{http://www.sdss3.org/surveys/boss.php}{\tt http://www.sdss3.org/surveys/boss.php}}
\citep{EW}
Wiggle\footnote{Dark Energy Survey :  \href{http://wigglez.swin.edu.au/}{\tt http://wigglez.swin.edu.au/}}
\citep{DJA}
DES\footnote{Dark Energy Survey: \href{http://www.darkenergysurvey.org/}{\tt http://www.darkenergysurvey.org/}}
\citep{DES} and (the forthcoming) 
EUCLID\footnote{EUCLID: \href{http://www.euclid-ec.org/}{\tt http://www.euclid-ec.org/}}
\citep{LAA}
In addition, the ongoing and future Cosmic Microwave Background (CMB) missions such as 
Planck\footnote{Planck: \href{http://www.cosmos.esa.int/web/planck/}{\tt  http://www.cosmos.esa.int/web/planck/}},
ACT\footnote{ACT: \href{http://www.physics.princeton.edu/act/}{\tt http://www.physics.princeton.edu/act/}},
and SPT\footnote{SPT: \href{http://pole.uchicago.edu/}{\tt http://pole.uchicago.edu/}}
surveys will map the CMB sky with unprecedented resolution.

The successful measurement of cosmological parameters relies on both the accuracy of the 
theoretical models as well as the precision of the statistics used. In the past, the precision of the
measurements was poor and a roughly ∼ 10\% statistical error on the measurement of the power spectrum
and even higher on the bispectrum was the limiting factor for discriminating among models and 
theories. However, current and forthcoming surveys 
are rapidly approaching the 1\% statistical precision for two-point statistics, and are constraining
higher-order statistics with similar level of improvement. This level of precision is comparable to the
accuracy of the theoretical models that have been developed.
In addition, the CMB sky at small angular scales is dominated by the secondaries, which are
highly non-Gaussian as they trace the underlying large-scale structure.
Consequently, a significant effort has been put into improving the theoretical development of new estimators
for gravity induced non-Gaussianity. These include the optimal estimators such as the skew-${\cal C}_\ell$
estimators \citep{skewspec}
or the kurt-${\cal C}_{\ell}$ estimators \citep{kurtspec}
as well as various sub-optimal morphological estimators \citep{MunshiWaerbeke}.

Analytical understanding of gravitational clustering is generally based on four different approaches: (1) Standard petrurbative analysis of
Euler-Continuity-Poisson system in the quasilinear regime \citep{review} in Eulerian framework (SPT)
or in Lagrangian space (LPT); (2) Physically motivated {\em ansatze} that capture
certain aspects of gravitational clustering in the non-linear regime \citep{BeS99}; (3) effective field theory (EFT)
based approaches \citep{EFT}; and (4) halo model and its variants \citep{halo}.

Gravity-induced higher-order correlation functions or their Fourier representations, the
higher-order polyspectra, can provide important clues to structure formation scenarios (see Ref.\cite{review} for a review).
Measurements of the power spectrum in a sub-volume of the survey is statistically correlated with
the average density contrast in that sub-volume. This correlation of this position
power spectrum and the average density-contrast was recently used
to define an estimator for the bispectrum  in the {\em squeezed-limit} \cite{komatsu}.
We will generalise the concept to position dependent power spectra to position-dependent
angular polyspectra and show how such constructions can be used as
estimators for higher-order polyspectra.  

Cumulant correlators (CCs) are natural generalisation of one-point cumulants and 
provide an alternative route to study higher order correlation hierarchy
and are well studied in the literature in the perturbative regime \cite{francis} and using hierarchical ansatz (HA) \cite{gen}. 
The Fourier representation of
the lower-order CCs i.e. the skew-spectrum (third order)\cite{skewspec} and
kurt-spectrum (fourth-order) \cite{kurtspec}
was also shown as an important form of data compression in 2D as well as in 3D.
We derive the cumulant correlators in the large separation limit and study their
relationship with the position-dependent multispectra hierarchy in the {\em soft} limit.

The organisation of the paper is as follows in \textsection\ref{sec:cc} we discuss the
Fourier transforms of the CCs; in \textsection\ref{subsec:quasi} and \textsection\ref{subsec:hier} we derive the
results for quasilinear and highly non-linear regime; the estimators for integrated bispectrum (IB)
and integrated trispectrum (IT) are described in \textsection\ref{sec:estimate};
the analytical expressions for bispectrum and trispectrum in squeezed limit are
presented in \textsection\ref{sec:unified_discussion} in a unified manner; 
in \textsection\ref{subsec:2D} we discuss the applications of these concepts to 2D (projected) surveys; 
the \textsection\ref{sec:res_disc}
is devoted to discussion of our results. We also present our
conclusions and point out the future prospects in this section. 
Finally, in Appendices-\textsection\ref{sec:tri}, \textsection\ref{sec:tri_collapsed}
and \textsection\ref{sec:tri_squeezed}
we extend the idea of IB to IT. 

We will concentrate on theoretical predictions in this paper. Comparison with numerical
simulations and extensions to popular halo model based approach will be presented in future work.
Observational aspects related to modelling of non-Gaussianities in CMB secondary maps
or issues related to galaxy redshift space distortions will also be dealt with elsewhere. 

For a discussion of the {\em soft} limits of polyspectra in the context of inflationary dynamics
(see \cite{inf1} and references there in). 
Certain aspects of the concept of polyspectra in the soft limit have been studied in the context of
large-scale structure formation \citep{valag1,valag2}; comparison against numerical
simulation was done in \citep{nishi1}. 

A note about our terminology is in order: by {\em polyspectra} we will mean the 
bispectrum, trispectrum and their higher-order analogs and with {\em multispectra}
will mean derived statistics e.g. skew-spectrum, kurt-spectrum or their higher-order versions (optimal
or sub-optimal).   
\section{Multispectra, Cumulant Correlators and the Large-Separation Limit}
\label{sec:cc}
The use of multispectra has become widespread recently.
The lowest order multispectrum (the skew-spectrum) probes the bispectrum \citep{skewspec}.
Its fourth-order analogues are the kurt-spectra which probe the trispectrum \citep{kurtspec}.
In the following we will establish the link between these multispectra and
their real-space analogs also known as the cumulant correlators \citep{francis}. This will allow us
to express the multispectra of all order in the limit of large wavenumber $k$. Our aim is to
elucidate the connection between the multispectra and the recently 
introduced integrated spectra.

The one-point cumulants $\la \delta_s^p({\bf x})\ra_c$ are collapsed multi-point correlation
functions when all the $p$ points are identified or collapsed to a single point;
see, e.g., Ref.\citep{review} for a review.
The cumulants are typically employed for study of
non-Gaussianity in many areas of cosmology including that of structure formation.
The subscript ``$_s$'' indicates smoothing of the density contrast $\delta({\bf x}) = (\rho({\bf x})-{\bar\rho})/{\bar \rho}$; 
where $\rho$ is the density at a point ${\bf x}$ and $\bar\rho$ is the average density $\bar\rho \equiv \la \rho({\bf x})\ra$
of the Universe smoothed using a suitable smoothing window. The {\em normalised} cumulants 
$S_p=\la\delta^p({\bf x}) \ra_c/\la\delta^2({\bf x})\ra_c^{p-1}$ are also used extensively in the literature;
see Ref.\citep{hypPT} for analytical estimates. 

The cumulant correlators (CC) are
natural generalisations of the one-point cumulants to two-point statistics $\la \delta_s^p({\bf x}_1)\delta_s^p({\bf x}_2)\ra_c$ 
\citep{francis,SzSz,MC00}. 
They are obtained by collapsing multipoint correlation functions of arbitrary order to two points. 
The normalised CCs denoted as ${\rm C}_{pq}$ are related to correlation function of order $(p+q)$ that are defined as \citep{francis}:
\ben
&& \la \delta_s^p({\bf x}_1) \delta_s^q({\bf x}_2) \ra_c \equiv {\rm C}_{pq}\, \sigma_s^{p+q-2}(R_0)\xi_{12}({x_{12}});\\
&& \vspace{1cm} \xi_{12}(x_{12})\equiv \la \delta_s({\bf x}_1)\delta_s({\bf x}_2) \ra_c; \quad x_{12} \equiv |{\bf x}_{12}| =|{\bf x}_1 -{\bf x}_2|.\\
&& \sigma^2_s \equiv \la \delta_s^2({\bf x}) \ra_c.
\een
For a concrete example, consider the lowest order in the hierarchy of CC, 
i.e. the two-to-one CC for a smoothed density contrast $\delta_i\equiv\delta_s({\bf x}_i)$.
We will be interested in the large separation limit  $x_{12}/R_0 \ll 1$.
This guarantees
that we have $\xi_{12}/\sigma^2_s(R_0)\ll 1$;  $\sigma^2_s(R_0)$ is the variance of the field obtained
using a top-hat smoothing window $W_{\rm TH}(kR_0)$ (to be defined below) of radius $R_0$ as:
\ben
S_{21}(x_{12}) \equiv \la\delta_1^2\delta_2\ra_c = \la \delta^2_s({\bf x}_1)\delta_s({\bf x}_2)\ra_c = {\rm C}_{21}\,\sigma_s^2\,\xi_{12}(x_{12});
\quad x_{12} = |{\bf x}_{12}|.
\label{eq:s21}
\een
Note we will use this form of smoothing throughout this paper.
Length scales which are in the perturbative regime ($\sigma^2(R_0)\ll 1$ where tree-level results
are valid) the normalised CCs typically become constant.
The CC $\la\delta_1^2\delta_2\ra_c$ is obtained by identifying two of the points involved in
a three-point 
correlation function $\la\delta_1\delta_2\delta_3\ra_c$, i.e ${\bf x}_1\equiv{\bf x}_3$. It retains
information
regarding the three-point correlation function from which it is derived but only for a {\em collapsed} configuration.
The Fourier-transform of Eq.(\ref{eq:s21}) also known as {\em skew-spectrum} $S_{21}(k)$ in the large-separation limit:
\ben
S_{21}(k^{\prime}) = \int {d^3 {\bf x}_{12} \over (2\pi)^3}\, S_{21}(x_{12})\exp[i {{\bf x_{12}}\cdot {\bf k}^{\prime}}]; 
\hspace{1cm} k^{\prime}=|{\bf k}^{\prime}|.
\een
We will use the wave-number ${\bf k}^{\prime}$ to represent the separation length-scale ${\bf x}_{12}$ and 
$\bk$ to denote the smoothing scale $R_0$ above in the Fourier domain.
We use the following expression  $S_{21}(x_{12})={\rm C}_{21}\sigma_s^2\xi(x_{12})$ valid 
in the large separation limit $x_{12} \rightarrow \infty$ i.e $R_0/x_{12}\ll 1$:
\ben
S_{21}(k^{\prime})= {\rm C}_{21}\,\sigma^2_s(R_0)\,P(k^{\prime}).
\een
The skew-spectrum in the Fourier domain, $S_{21}(k)$, represents the bispectrum in the squeezed limit.
In general ${\rm C}_{21}$ is not a constant but a function of smoothing radius $R_0$, or
equivalently the length scale $k$.
The power spectrum is defined through the Fourier-transform of the correlation function $\xi_s$: 
\ben
P(k^{\prime}) = \int {d^3 {\bf x}_{12} \over (2\pi)^3}\, \xi_s(x_{12})\exp[i {{\bf x_{12}}\cdot {\bf k}^{\prime}}].
\een
The higher-order cumulant correlators $S_{pq}(x_{12})$ are natural generalisations of the two-to-one cumulant
correlator defined above:
\ben
S_{pq}(x_{12}) \equiv \la\delta_1^p\delta_2^q \ra_c = \langle\delta^p({\bf x}_1)\delta^q({\bf x}_2) \rangle_c 
= {\rm C}_{pq}\, \xi_{12}(x_{12})\, \sigma_s^{p+q-2}(R_0).
\label{eq:spq}
\een
The corresponding Fourier-transform defines the related collapsed multispectra $S_{pq}(k)$:
\ben
S_{pq}(k^{\prime}) = \int {d^3 {\bf k} \over (2\pi)^3}S_{pq}(x_{12})\exp[i {{\bf x_{12}}\cdot {\bf k}^{\prime}}].
\label{eq:spq_k}
\een
Using Eq.(\ref{eq:spq}) in Eq.(\ref{eq:spq_k}) we arrive at the following expression:
\ben
S_{pq}(k^{\prime}) = {\rm C}_{pq}\, \sigma_s^{p+q-2}(R_0)\, P(k^{\prime}). \label{eq:most_important_result}
\een
The expressions for the lower order ${\rm C}_{pq}$ are given below in Eq.(\ref{eq:c21}). 
Eq.(\ref{eq:most_important_result}) is the one of the important result of this paper. 
We will see that the position-dependent spectra we consider later in
this paper have a structural similarity to the expressions for multispectra derived
above in the above limit. We shall show that, for the bispectrum in the squeezed limit,
the results are formally identical to the skew-spectrum at low $k$ limit,
though the mathematical interpretation is different.   
The normalised CC or ${\rm C}_{pq}$ are in general functions of the smoothing scale $R_0$ (equivalently the wavenumber $k$).
The $k$ dependence manifests itself as logarithmic slope $n$ dependence of the power spectrum.

\begin{figure}
\vspace{1.25cm}
\begin{center}
{\epsfxsize=13. cm \epsfysize=5. cm 
{\epsfbox[32 411 547 585]{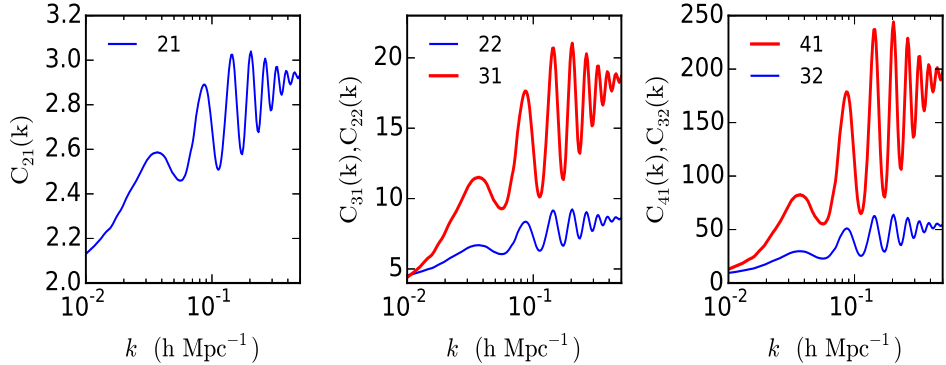}}}
\caption{The 3D normalised cumulant corelators [defined in Eq.(\ref{eq:c21})-Eq.(\ref{eq:c31})]
are plotted. The plots show 
${\rm C}_{21}(k)$ (left panel), ${\rm C}_{31}(k)$ and ${\rm C}_{22}(k)$ 
(middle panel)
and ${\rm C}_{41}(k)$ and ${\rm C}_{32}(k)$ (right panel) as a function of the $k$ wave number associated
with the inverse of the radius of the top-hat smoothing window $R_0$.
The results are derived using Standard Perturbation Theory (SPT) and 
the effective spectral index $n$ was computed using the linear power spectrum.}
\label{fig:cumu1}
\end{center}
\end{figure}

The lower-order CCs are plotted in Figure-\ref{fig:cumu1} as functions of $k\, (h^{-1}\rm Mpc)$.
We plot ${\rm C}_{21}$ (left-panel) ${\rm C}_{31}$ and 
${\rm C}_{22}$ (middle-panel) and ${\rm C}_{41}$ and ${\rm C}_{32}$
(right-panel). The oscillations correspond to BAO signature in the underlying power spectrum.
These plots depict the asymptotic value of the multispectra
in the limit $k^{\prime}\rightarrow 0$ as a function of $k$. In this limit the normalised CCs
or ${\rm C}_{pq}$ are independent of $k^{\prime}$ and the $k^{\prime}$ dependence of $S_{pq}$ is
completely absorbed in $P(k^{\prime})$. The ${\rm C}_{pq}$
are functions of local slope of the power spectrum.
\section{Quasilinear Regime: Tree-level Results in the Soft (Squeezed) Limit}
\label{subsec:quasi}
The two-point (joint) probability distribution function (PDF) for the smoothed (using a top-hat window) density field $\delta_s$
can be expressed in terms of the one-point $p_{\delta}(\delta)$, bias $b_{\delta}(\delta)$ in the large 
separation limit ${\xi_{12}/\sigma_2^2} \ll 1$. Such a limiting situation is reached when the two cells are
separated by a distance relatively larger than the smoothing scale. 
\ben
p_{\delta}(\delta_1,\delta_2)d\delta_1d\delta_2 = p_{\delta}(\delta_1)p_{\delta}(\delta_2)
[1 + b_{\delta}(\delta_1)\,\xi^{\delta\delta}_{12}b_{\delta}\,(\delta_2)]d\delta_1 d\delta_2
\label{eq:2pdf}
\een
The CCs introduced in \textsection{\ref{sec:cc}} are normalised two-point moments $\la \delta_1^p \delta_2^q \ra_c$ and can be expressed as:
\ben
&& {\rm C}^{\delta\delta}_{pq} \equiv{\la\delta_1^p\delta_2^q\ra_c \over \la\delta^2\ra_c^{p+q-2}\la\delta_1\delta_2\ra_c}; \quad
\la\delta_1^p\delta_2^q \ra_c = \int_{-1}^{\infty}\int_{-1}^{\infty} \delta_1^p\,\delta_2^q \;p(\delta_1,\delta_2) d\delta_1d\delta_2.
\een
Normalisation requires 
$\int_{-1}^{\infty} d\delta_1\,\int_{-1}^{\infty}\,d\delta_2\, p_{\delta}(\delta_1,\delta_2)=1$ 
and $\int^{\infty}_{-1} d\delta\,p_{\delta}(\delta)=1$ giving us the constraint ${\rm C}^{\delta\delta}_{11}=1$. 
In the large-separation limit the following factorisation property holds:
\ben
{\rm C}^{\delta\delta}_{pq} = {\rm C}^{\delta\delta}_{p1}{\rm C}^{\delta\delta}_{q1}.
\label{eq:factor}
\een
In the quasilinear (perturbative) regime, the leading order terms of the entire hierarchy of ${\rm C}_{pq}$
can be evaluated analytically \cite{francis}. 
We quote here the following lower-order expressions:
\ben
\label{eq:c21}
&& {\rm C}^{\delta\delta}_{21}= {68 \over 21} +{\gamma_1 \over 3}; \\
&& {\rm C}^{\delta\delta}_{31}= {11710 \over 441} + {61 \over 7}\gamma_1 + {2 \over 3}\gamma_1^2 + {\gamma_2 \over 3}; \label{eq:c31}
\een
where the factors $\gamma_p$ are defined as follows:
\ben
\gamma_p =  {{\rm d}^p {\rm log} \sigma^2(R_0) \over {\rm d} ({\rm log} R_0)^p}.
\een
These results ignore contributions from loop diagrams and are thus valid only in the limiting situation when $\la\delta^2\ra\ll 1$.
For power-law power spectra $P(k) \propto k^n$ we have $\gamma_1= -(n+3)$ and $\gamma_p=0$ for $p>1$.
In this limit the coefficients are polynomials in $n+3$, a property they share with the 
integrated spectra that we will study later.
In case of the skew-spectra, the lowest order polynomial (i.e. linear) in this family, the coefficients
match with those of the integrated bispectra (to be defined later) but this is not the case for higher order spectra.
This is also true for the divergence of velocity $\Theta$.
For $n=-3$ these results represent statistics of unsmoothed fields and their values are determined
completely by the angular averages of the tree-level amplitudes $\nu_n$. In this limit they
can be analysed by the HA (see \textsection\ref{subsec:hier}).

We will use the concept of ${\rm C}_{pq}$ for the case of velocity divergence 
$\Theta = -\nabla\cdot{\bf v}/H$
(to be introduced and discussed in more detail in \textsection\ref{sec:unified_discussion}) 
and generalise the concept of the integrated bispectrum to $\Theta$. 
It is possible to consider mixed cumulant correlators of $\delta$ and $\Theta$ e.g. $\la\delta_1^p\Theta_2^q \ra$.
In this case following similar arguments we can write:
\ben
&& {\rm C}^{\delta\Theta}_{pq} = {\la\delta_1^{p}\Theta_2^q\ra_c \over \la\delta_1^2\ra^{p-1} \la\Theta_2^2\ra^{q-1}\la\delta_1\Theta_2\ra};\\
&& {\rm C}^{\delta\Theta}_{pq} = {\rm C}^{\delta\delta}_{p1}{\rm C}^{\Theta\Theta}_{q1}
\label{eq:factor_pdf_bias}
\een
The corresponding joint PDF that generalises Eq.(\ref{eq:2pdf}) is given by:
\ben
p_{\delta\Theta}(\delta_1,\Theta_2)d\delta_1d\Theta_2 = p_{\delta}(\delta_1)p_{\Theta}(\Theta_2)
[1 + b_{\delta}(\delta_1)\,\xi^{\delta\Theta}_{12}\,b_{\Theta}(\Theta_2)]d\delta_1 d\Theta_2.
\label{eq:bias}
\een
Here, $p_{\delta\Theta}$ is the joint PDF for $\delta_1\equiv\delta({\bf x}_1)$ and $\Theta_2\equiv\Theta({\bf x}_2)$.
The one-point PDFs for $\delta$ and $\Theta$ are denoted as $p_{\delta}(\delta)$ and $p_{\Theta}(\Theta)$.
the corresponding bias functions are defined as $b_{\delta}$ and $b_{\Theta}$ respectively.
The correlation function of $\delta$ and $\Theta$ is denoted $\xi_{12}^{\delta\Theta}\equiv \la\delta_1\Theta_2 \ra$.
In the Fourier domain we can similarly define mixed multispectra and their squeezed limits which
can provide consistency checks on results obtained using $\delta$ and $\Theta$ fields alone.   
\section{Highly Non-linear Regime: Hierarchical {\em ansatz} (HA) in the Soft Limit}
\label{subsec:hier}
Gravity is scale-free. In the absence of of an externally-imposed length scale, such as might be set by
initial conditions, it is reasonable to assume that gravitational clustering should evolve towards
a scale-invariant form, at least on small scales where gravitational effect dominates
over initial conditions \citep{DP77,GP77,Fry84,Ber92,Ber94}. Observations offer
support for such an idea, in that the observed two-point correlation function $\xi_2(x)$ of 
galaxies is reasonably well represented by a power law over quite a large range of length scales,
$\xi_2(r) \equiv \left( {r/ 5h^{-1}\rm Mpc}\right )^{-\gamma}$
between $100h^{-1}$ {\rm kpc} and $10 h^{-1}${\rm Mpc}. Higher-order correlation
functions of galaxies also appear to satisfy a scale-invariant form, with $\xi_{\rm N} \propto \xi_2^{\rm N-1}$
as expected from the application of a
general scaling {\it ansatz} 
\citep{GP77,FryPeebles78,DP77}

For example, the observed lower-order correlation function exhibits a hierarchical form
\ben
\label{eq:two}
&& \xi_{ab} \equiv \xi_2({\bf x}_a, {\bf x}_b); \\
&& \xi_{abc}\equiv \xi_3({\bf x}_a,{\bf x}_b,{\bf x}_c) \equiv \la\delta({\bf x}_a)\delta({\bf x}_b)\delta({\bf x}_c)\ra_c
= Q_3(\xi_{ab}\xi_{bc}+ \xi_{bc}\xi_{ca}+\xi_{ab}\xi_{ac}) \label{eq:three};\\
&& \xi_{abcd} \equiv \xi_4({\bf x}_a,{\bf x}_b,{\bf x}_c,{\bf x}_d) \equiv \la\delta({\bf x}_a)\cdots\delta({\bf x}_d)\ra_c \nn \\
&& \hspace{1.cm}=R_a(\xi_{ab}\xi_{bc}\xi_{cd}+{\rm cyc.perm.}) + R_b(\xi_{ab}\xi_{ac}\xi_{ad}+{\rm cyc.perm.}); \label{eq:four} \\
&& \xi_{abcde} \equiv \xi_5({\bf x}_a,\cdots,{\bf x}_e) \equiv \la\delta({\bf x}_a)\cdots\delta({\bf x}_e)\ra_c \nn\\
&& \hspace{1.7cm}=S_a(\xi_{ab}\xi_{bc}\xi_{cd}\xi_{de}+{\rm cyc.perm.}) + S_b(\xi_{ab}\xi_{bc}\xi_{bd}\xi_{de}+{\rm cyc.perm.}) \nn \\
&& \hspace{1.7cm} + S_c(\xi_{ab}\xi_{ac}\xi_{ad}\xi_{ae}+ {\rm cyc.perm.} ). \label{eq:five}
\een
The hierarchy of equations - the
Born, Bogolubov, Green, Kirkwood, Yvon (BBGKY) 
hierarchy that governs the evolution of the $p$-body density functions (in the full
phase space) has been established
for matter in an expanding universe  \citep{Peebles80}. 
Although the exact nature of this correlation hierarchy can only be obtained 
by solving the full set of BBGKY equations.
The exact nature of this correlation hierarchy can only be understood by solving the full
set of BBGKY equations, which in general can not be done \citep{DP77,GP77,Fry84}.

Useful insights can nevertheless be obtained by investigating the consequences of scaling 
properties to general closure \citep{BaS99,newHam88b} schemes based fact that the hierarchy admits 
self-similar solutions \citep{DP77}. The evolution of the power spectrum has also been
tackled in a similar way \citep{HKLM91}. In this approach the higher-order correlation
functions can be expressed as:
\ben
\xi_{\rm N}({\bf x}_1,\cdots, {\bf x}_{\rm N})= \sum_{\alpha,{\rm N-trees}} Q_{N,\alpha} \sum_{\rm labelling}\prod^{{\rm N}-1}_{\rm edges} 
\xi_2({\bf x}_i,{\bf x}_j).
\label{eq:hier}
\een
Note that there are no theoretical predictions for the topological amplitudes $Q_{N,\alpha}$ 
in this approach. Perturbative calculations have shown that gravity can induce  a similar
hierarchy starting from Gaussian initial conditions \citep{Fry84,Ber92,Ber94}
in the limit of weak clustering.

This {\em tree-level} model of hierarchical clustering however is a particular case of a more general scaling ansatz proposed by
\citep{BaS99}, in which the N –point correlation functions can be written in the form
\ben
\xi_{\rm N}(\lambda {\bf x}_1, \cdots, \lambda {\bf x}_{\rm N})= \lambda^{{\rm N}-1} \xi_{\rm N}({\bf x}_1,\cdots, {\bf x}_{\rm N})
\een
See, e.g., Ref.\citep{BeS99} and the reference therein. We shall work with the minimal hierarchical models
as they distil some very basic features shared by other more complicated models.
In the Fourier domain the equivalent results relate the higher-order polyspectra with the ordinary power sepctrum \citep{MC00}.
The bispectrum can be obtained by taking the Fourier transform of Eq.(\ref{eq:three}):
\ben
&& \la\delta({\bf k}_1)\delta({\bf k}_2)\delta({\bf k}_3)\ra_c\ 
\equiv (2\pi)^3\,\delta_{\rm D}({\bf k}_{123})\,B_2({\bf k}_1, {\bf k}_2, {\bf k}_3)  ;\\
&& B_2({\bf k}_1, {\bf k}_2, {\bf k}_3) = 
Q_3 \left[ P({k}_1)P({k}_2)+ P({k}_2)P({k}_3)+P({k}_1)P({k}_3)\right ].
\een
Throughout we will use ${\bf k}_{12\cdots p}={\bf k}_1 + {\bf k}_2 +\cdots + {\bf k}_p$.
The trispectrum $B_3({\bf k}_1,\cdots,{\bf k}_4)$ is expressed in terms of two hierarchical amplitudes,
$R_a$ and $R_b$, introduced in Eq.(\ref{eq:three}):
\ben
&& \la\delta({\bf k}_1)\cdots\delta({\bf k}_4)\ra_c \equiv (2\pi)^3 \delta_D({\bf k}_{1234}) B_3({\bf k}_1,\cdots, {\bf k}_4)\\
&& B_3({\bf k}_1,\cdots, {\bf k}_4) =  R_a \left [ P({k}_1)P(|{\bf k}_{12}|)P(|{\bf k}_{123}|) + {\rm cyc.perm.} \right ]\nn \\
&& \hspace{3cm}+ R_b [P(k_1)P(k_2)P(k_3) + {\rm cyc.perm.}].
\een
The next-order multispectrum $B_4({\bf k_1,\cdots,k_5})$ is obtained by taking FT of Eq.(\ref{eq:four}):
\ben
&& \la\delta({\bf k}_1)\cdots\delta({\bf k}_5)\ra_c \equiv (2\pi)^3 \delta_D({\bf k}_{1234}) B_4({\bf k}_1,\cdots, {\bf k}_5)\\
&& B_4({\bf k_1,\cdots,k_5}) = S_a \Big [ P({k}_1)P({|\bf k}_{12}|)P({|\bf k}_{123}|)P(|{\bf k}_{1234}|)+{\rm cyc.perm.} \Big ] \nn \\
&& \hspace{3cm}+ S_b \left [ P({k}_1)P({k}_2)P(|{\bf k}_{123}|)P(|{\bf k}_{1234}|) + {\rm cyc.perm.} \right ] \nn \\
&& \hspace{3cm}+ S_c \left [ P({k}_1)P({k}_2)P({k_3})P({k_4}) + {\rm cyc.perm.} \right ].  
\een
The result presented in Eq.(\ref{eq:most_important_result}) is derived using very general arguments. 
In the rest of this Section we will work out in detail for few specific models.

In the highly non-linear regime the higher-order correlation functions can be calculated
using a hierarchical ansatz (HA) \cite{gen}.
The parameters $\{ Q_3 \}$, $\{ R_a,R_b \}$ and $\{S_a,S_b,S_c\}$ are topological amplitudes of various 
tree diagrams 
used to represent the correlation hierarchy at third fourth and fifth order, respectively. 
For specific models see Ref.\cite{BaS99,BeS92,BeS99,SzSz}.
The lower-order linear combinations of these amplitudes that produce the one-point cumulants or $S_N$
have been studied using numerical simulations \cite{hypPT}.

In our calculation we will take the specific model by Bernardeau \& Schaeffer \cite{BeS92} where we identify
$Q_3=\nu_2$, $R_a=\nu_2^2, R_b=\nu_3$ and $S_a=\nu_4, S_b=\nu_3\nu_2, S_c=\nu_2^3$.
In the model proposed by Szapudi \& Szalay \citep{SzSz} the tree amplitudes of a given
order have identical values: $R_a=R_b$ and $S_a = S_b = S_c$ or in general in Eq.(\ref{eq:hier}) $Q_{N,\alpha}=Q_N$.

In the quasilinear regime the vertices develop angular dependence on the wave vectors
${\bk}_i$. In the tree-level perturbative regime the same tree hierarchy can be used
and in the absence of smoothing the angular averaged biases can replace the
corresponding $\nu_n$ s \citep{Fry84,Ber92}. The power spectrum in the quasilinear regime is 
replaced by the linear power spectrum $P_{\rm L}(k)$. This is the regime we will
use in this paper. We will omit the subscript $_L$ henceforth.  
\subsection{Bispectrum in the soft limit}
\label{subsec:bi}
The influence of large-scale density fluctuations on structure formation results in the
coupling of small and large-scale modes. At the lowest order such coupling
can be described by the corresponding bispectrum in the so-called ``squeezed'' configuration.
In the squeezed limit one of the wavenumbers, $k_1$, of the triangle representing the bispectrum
in the Fourier domain, is much smaller than the other two i.e. $k_1\ll k_2\approx k_3$,
thus, as we will see, effectively reducing the bispectrum to a power spectrum.
In this limit the following parametrization applies:
\ben
&& B_2({\bf k-q_1, -k+q_{12}, -q_2}) = 
Q_3 \Big[ P(|{\bf k-q_1}|)P(|{\bf -k+q_{12}}|) \nn \\
&& \hspace{3cm} + P(|{\bf -k+q_{12}}|)P(q_2)+ 
P(|{\bf k-q_2}|)P({q}_2)\Big ].
\label{eq:b2_all_k}
\een
In our derivation, we will expand the power spectra in a Taylor-series as follows:
\ben
\label{eq:taylor1}
&& P(|{\bf k-q_1}|) = P(k) \left [1 - {{\bf k \cdot q_1}\over k^2} {d\ln P(k) \over d\ln k} +\cdots \right ];\\
&& P(|{\bf -k+q_{12}}|) = P(k) \left [1 - {{\bf k \cdot q_{12}}\over k^2} {d\ln P(k) \over d\ln k} +\cdots \right ].
\label{eq:taylor2}
\een
Unlike the perturbative bispectrum the hierarchical bispectrum does not display any Infrared (IR) 
divergence. In the squeezed limit $k\gg q_3$, so we ignore the terms of ${\cal O}(q_i/k)$ 
so that Eq.(\ref{eq:b2_all_k}) takes the following form:
\ben
&& B_2({\bf k-q_1, -k+q_{12}, -q_2}) \stackrel{\text{squeeze}}{\approx} Q_3[2P(q_2)P(k)+P^2(k)]; \nn \\
&& \lim_{\bq_2 \rightarrow 0} B_2({\bf k, -k, -q_2})  \stackrel{\text{squeeze}}\approx 2Q_3  P_{L}(q_2)P(k).
\label{eq:b2_first_k}
\een
The corrections from the Taylor expansion in Eqs.(\ref{eq:taylor1}-\ref{eq:taylor2}) are only of  ${\cal O}(q_i/k)^2$.
Notice that we have also ignored terms of ${\cal O}[P(q_i)/P(k)]$ for CDM-like 
spectrum for  $(q_i/k)\ll 1$.
The subscript $_L$ denotes the linear power spectrum. The power spectrum is effectively in the linear regime
for long wavemodes.
This matches with the expression in Eq.(\ref{eq:q3}). In the last term we have assumed for a CDM like
spectrum $P(k)\ll P(q_2)$ for $k\gg q_3$. This is consistent with the result obtained in real space \cite{MC00}:
\ben
&& \la\delta_1^2\delta_2 \ra_c = {\rm C}_{21} \xi_{12}\sigma_L^2 ; \quad {\rm C}_{21} =  2Q_3. \label{eq:q3}
\een
The real space result can be obtained by identifying two of the points involved in a three-points $a=b$
and demanding $\xi_{ac}=\xi_{bc} \ll \xi_{ab}$ in Eq.(\ref{eq:three}) to neglect the
linear order terms in $\xi_{ac}/\xi_{aa}$ ($\xi_{aa}\equiv\sigma_L^2$).

In the specific model of Bernardeau \& Schaeffer $Q_3=\nu_2$. In the perturbative regime the
the unsmoothed results can be reproduced by taking $n=-3$ which gives $\nu_2 = {34/21}$.
Using this result we reproduce the result by Bernardeau in Ref.\cite{francis}, i.e. ${\rm C}_{21}= {68/21}$. 
\subsection{Trispectrum in the soft limit}
\label{subsec:tri}
In the soft limit the trispectrum can take  either a {\em squeezed} or {\em collapsed} shape. 
In the squeezed case we have a configuration in which the trispectrum
has one side  much smaller than the others. In this configuration the
trispectrum can be described effectively as a product of the bispectrum $B_2(\bk_a,\bk_b,\bk_c)$ and the power spectrum $P(q)$; here $\bq$ is the ``soft'' mode. We will use the following parametrization:
\ben
&& B_3({\bf k_a-q_1, k_b-q_2, k_c+q_{123}, -q_3}) \stackrel{\text{squeeze}}{=}  \nn \\
&& R_a [ P(q_3)\left \{ P(k_a)P(k_b) + {\rm cyc.perm.} \right \} +P(k_a)P(k_b)P(k_c) ] \nn \\
&& +R_b[2P(q_3)\{ P(k_a)P(k_b) +{\rm cyc.perm.} \} + \{ P(k_a)[P^2(k_b)+P^2(k_c)] + {\rm cyc.perm.} \} ]\nn \\
\label{eq:b3_first_k}
\een
In the limit $k_a,k_b,k_c \ll q_3$ we have $P(k_a),P(k_b),P(k_c)\gg P(q_3)$, so the terms that survive are:
\ben
&& \lim_{\bq_i\rightarrow 0}B_3({\bf k_a-q_1, k_b-q_2, k_c+q_{123}, -q_3}) \stackrel{\text{squeeze}}{\approx} 
\lim_{\bq_3\rightarrow 0}B_3({\bf k_a, k_b, k_c, -q_3})   \nn \\
&& \quad\quad \approx (R_a + 2R_b)P(q_3)[ P(k_a)P(k_b) + {\rm cyc.perm.} ] \delta_{\rm D}({{\bf k}_{abc}}).
\label{eq:b3_second_k}
\een
Both ``snake'' and ``star'' terms contribute to the trispectrum in the squeezed limit.
The effective bispectrum that describes the trispectrum in the squeezed limit has an amplitude $(R_a + 2R_b)$
rather than $Q_3$.
For the other soft configuration we consider the case when one of the diameter of the quadrilateral
representing the trispectrum is much smaller compared to its sides, also known as the collapsed configuration.
In this configuration only the ``snake'' terms contribute: 
\ben
&& \lim_{\bq \rightarrow 0} B_3({\bf k_1, -k_1-q, k_2, -k_2+q})  
\stackrel{\text{collapsed}}{=} 2R_b \Big [ 2\, P(k_1)P(k_2)P(q) + 2\, P(k_1)P(|{\bf k_{12}}|)P(k_2) \nn \\
&& \hspace{3cm}+ \left [ P^2(k_1)+P^2(k_2) \right ] P(|{\bf k_{12}}|) \Big ]; \\ 
&& B_3({\bf k_1, -k_1, k_2, -k_2}) \stackrel{\text{collapsed}}{\approx} 4R_b\, P(k_1)P(k_2)P(q).
\label{eq:squeezed2}
\een
In the collapsed configuration the trispectrum reduces to a product of three power spectra.
The Fourier-space expressions
in Eq.(\ref{eq:b3_first_k}) and Eq.(\ref{eq:b3_second_k}) correspond respectively to Eq.(\ref{eq:b3_first}) and Eq.(\ref{eq:b3_second})
in real-space \citep{MC00}:
\ben
&& \la\delta_1^3\delta_2 \ra_c={\rm C}_{31}\xi_{12}\sigma_L^4; \quad  {\rm C}_{31} = (3R_a + 6R_b); \label{eq:b3_first}\\
&& \la\delta_1^2\delta_2^2 \ra_c={\rm C}_{22}\xi_{12}\sigma_L^4; \quad {\rm C}_{22} = 4R_b. \label{eq:b3_second}
\een
Joint measurements of ${\rm C}_{31}$ and ${\rm C}_{22}$ can be used to estimate the amplitudes $R_a$ and $R_b$:
if we use $R_a\equiv\nu_3=682/189$ and $R_b \equiv \nu_2^2 =(34/21)^2$ we recover
the result in \citep{francis} ${\rm C}_{31}={11710/441}$ and ${\rm C}_{22}= {(68/21)^2}$.
\begin{figure}
\vspace{1.25cm}
\begin{center}
{\epsfxsize=13. cm \epsfysize=5. cm 
{\epsfbox[32 411 547 585]{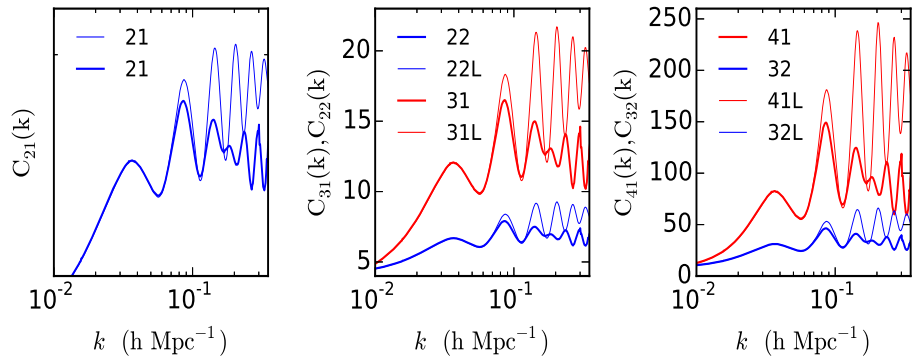}}}
\caption{The 3D normalised cumulant corelators [defined in Eq.(\ref{eq:c21})-Eq.(\ref{eq:c31})]
are plotted. The plots show 
${\rm C}_{21}(k)$ (left panel), ${\rm C}_{31}(k)$ and ${\rm C}_{22}(k)$ (middle panel)
and ${\rm C}_{41}(k)$ and ${\rm C}_{32}(k)$ (right panel) as a function of the $k$ wave number.
The results are derived using a standard perturbation theory (SPT) and power spectrum including one-loop
corrections. The results shown are for z=0 (see text for more details).}
\label{fig:cumu2}
\end{center}
\end{figure}

Previous studies have focused on many different aspects of such theories,
including one-point probability distribution, the void-probability distribution function
and joint probability distribution function \citep{BaS99,BeS92,BeS99} which are directly related to the bias of 
over-dense objects \citep{bias}. Multi-point correlation function, cumulants and cumulant correlators 
of over-dense objects to arbitrary order have also been considered \citep{gen,snst1,snst2,scaling}. 
The results presented here extend these results into the Fourier domain.
We show how squeezed configurations of polyspectra of arbitrary order 
can be studied by using local estimates of lower order polyspectra.
\section{Estimators for Polyspectra in their Soft Limit}
\label{sec:estimate}
In this Section we will develop a theory of the estimators for the squeezed multispectra.
We will consider a density field $\delta({\bf r})$ defined in a simulation box of side $L_{\rm box}$. We will also consider $N^3$
identical cubic sub-volumes  with sides of length $L=L_{\rm box}/N$. The cosmological statistics measured in a sub-volume centred at the
position $r_L$ will be denoted $L$; the volume will be denoted $V_L=L^3$.
To compute the squeezed higher-order multispectra
we will cross-correlate the statistics measured in the entire simulation box against those estimated from
these sub-volume. We will consider 3D surveys in this section but a generalisation to projected or 2D survey
will be dealt with in \textsection\ref{subsec:2D}.
The results we present can be generalised to the case of observational data with minimal changes.

The local mean-density perturbations relative to the global mean density of the main volume is denoted as ${\bar\delta}({\bf r}_L)$
and can be expressed through the following convolution:
\ben
\bar \delta({\bf r}_L) 
= {1 \over V_L}\, \int\, d^3{\bf r}\, \delta({\bf r})\,W_L({{\bf r}-{\bf r}_L}).
\een
The window function defined as $W_L({\bf x}) \equiv \prod_{i=1}^{i=3}\theta({x}_i)$. The one-dimensional unit step
functions satisfy $\theta({x}_i)=1$ for $x_i \le L/2$ and zero otherwise. The equivalent expression
in the Fourier domain takes the following form:
\ben
&& \bar \delta({\bf k},{\bf r}_L)=
 \int {d^3{\bf q}\over (2\pi)^3}\; \delta({\bf k}-{\bf q})\; W_L({\bf r}-{\bf r}_L)\;\exp(-i{\bf r}\cdot{\bf k}).
\een
The window $W_L$ in the Fourier domain is given by:
\ben
W_L({\bf q}) \equiv V_L \prod_i^{3} {1 \over q_i}{{\rm sinc}({q_i\,L\over 2})};
\een
where ${\rm sinc}(x) = \sin(x)/x$. The window has the following property which we will use throughout in our derivation:
\ben
W_L^2({\bf r})=W_L({\bf r}); \quad W_L({\bf q}_1) =  \int {d^3{\bf q}_2  \over (2\pi)^3} W_L({\bf q}_2)W_L({\bf -q_{12}}).
\label{eq:window}
\een
 The position dependent power spectrum
$P({\bf k}; {\bf r}_L) \equiv |\delta({\bf k}; {\bf r}_L)|^2/ V_L$
estimated from a sub-volume is given by the following expression:
\ben
P({\bf k}; {\bf r}_L)  =
{1 \over V_L}\int {d^3 {\bf q}_1 \over (2\pi)^3}\int {d^3 {\bf q}_2 \over (2\pi)^3}
\delta({\bf k}-{\bf q}_1 )\delta(-{\bf k}-{\bf q}_2 )W_L ({\bf q}_1 )W_L ({\bf q}_2 ).
\label{eq:local_ps}
\een
This estimate of the local power spectrum can now be used to construct estimators for 
bispectrum and trispectrum in the soft limit.
\subsection{Estimator of the Squeezed Bispectrum}
\label{subsec:bi_est}
The squeezed bispectrum can be estimated by cross-correlating the local estimates of the
density contrast and the local power spectrum \citep{komatsu}:
\ben
&&  \la P(k)\bar\delta({\bf r}_L)\ra_c = \quad {1 \over V_L^2}\int {d^3{\bf q}_1 \over (2\pi)}\dots\int {d^3{\bf q}_3 \over (2\pi)}
\la \delta({\bf k-q_1})\delta({\bf -k-q_2})\delta(-{\bf q}_3) \ra \nn \\
&& \quad\quad \times W_L({\bf q}_1)W_L({\bf q}_2)W_L({\bf q}_3)\delta_{\rm 3D}({\bf q}_{123}).
\een
Using Dirac $\delta_{3D}$ function to reduce the dimensionality of the above integral gives
\ben
&& \la P({k}) \bar\delta({\bf r}_L)\ra_c= \quad {1 \over V_L^2}\int {d^3{\bf q}_1 \over (2\pi)^3}
\int {d^3{\bf q}_2\over (2\pi)^3}B_2[{\bf k-q_1, -k+q_{12}, -q_2}]\nn \\
&& \hspace{3cm} \times W_L({\bf q}_1)W_L(-{\bf q}_{12})W_L(-{\bf q}_2).\\
&& \la P({k}) \bar\delta({\bf r}_L)\ra_c = 2Q_3 \sigma_L^2 P(k).
\label{eq:final_bispec}
\een
The derivation uses the result in Eq.(\ref{eq:b2_first_k}).

In general the vertex $Q_3$ is defined in the Fourier space and carries an angular dependence.
Integrating out this dependence gives the integrated bispectrum $\bar{\cal B}_{21}$:
\ben
{\cal B}_{21}({\bf k},r_L)=\langle P({\bf k},r_L) {\bar\delta}(r_L) \rangle_c; \quad
\bar{\cal B}_{21}({k}) \equiv \int{ d\hat\Omega_{k} \over 4\pi}{\cal B}_{21}({\bf k},r_L).
\label{eq:define_ibispec}
\een
\subsection{Estimators of Trispectrum: Squeezed and Collapsed}
\label{subsec:tri_est}
As we have previously mentioned, in the soft limit the trispectrum $B_3$ exists in {\em squeezed} and {\em collapsed} configuration, which we discuss next. 
We will show that trispectrum in the squeezed limit can be constructed by correlating the local estimates
of the bispectrum $B_2$ and the local average density contrast $\bar\delta$. The collapsed limit of
the trispectrum is constructed using covariance matrix for the local power spectrum.

{\bf Squeezed:} Local estimates of the bispectrum from a small patch of a survey
and the average density contrast
measured from the same patch are correlated. The correlation is a measure of 
the trispectrum in the squeezed limit described in Eq.(\ref{eq:b3_first_k}):
\ben
&&  \la B_2 \bar\delta({\bf r}_L)\ra_c \equiv \la B_2({\bf k_a,k_b,k_c};r_L)\bar\delta({\bf r}_L)\ra_c  \nn \\
&& = \quad {1 \over V_L^2}\int {d^3{\bf q}_1 \over (2\pi)}\dots\int {d^3{\bf q}_4 \over (2\pi)}
\la \delta({\bf k_a-q_1})\delta({\bf k_b-q_2})\delta({\bf k_c-q_3})\delta(-{\bf q}_4) \ra \nn \\
&& \quad\quad \times W_L({\bf q}_1)W_L({\bf q}_2)W_L({\bf q}_3)W_L({\bf q}_4)\,\delta_{\rm D}({\bf q}_{1234})\delta_{\rm D}({\bf k}_{abc}).
\een
Integrating out the variable ${\bf q}_4$ collapses the above 4D integral to a 3D integral: 
\ben
&& \la B_2 \bar\delta({\bf r}_L)\ra_c= \quad {1 \over V_L^2}\int {d^3{\bf q}_1 \over (2\pi)^3} 
\cdots \int {d^3{\bf q}_3\over (2\pi)^3} B_3[{\bf k_a-q_1,k_b-q_2,k_c+q_{123},-q_3}]\nn \\
 &&\hspace{3cm} W_L({\bf q}_1)W_L({\bf q}_2)W_L(-{\bf q}_{123})W_L({\bf q}_3).\\
&& {\cal T}_{31}(\bk_a,\bk_b,\bk_c) \equiv \la B_2 \bar\delta({\bf r}_L)\ra_c = (R_a+2 R_b)\sigma_L^2 [ P(k_a)P(k_b) + {\rm cyc.perm.} ].
\label{eq:final_bispec1}
\een
We have used the expression in Eq.(\ref{eq:b3_first_k}) for our derivation.

{\bf Collapsed:} For the other ``soft'' configuration the sides of the quadrangle are much bigger compared to one of its diagonal.
We have ignored the terms that are of ${\cal O}({{q}/{k}_i})$. This is the Fourier analogue of the expression in Eq.(\ref{eq:b3_second}):

\ben
&& \la P({\bf k}_a,{\bf r}_L)P({\bf k}_b,{\bf r}_L) \ra_c = \delta_D(\bk_{12}) {1 \over V^2_L}\int {d^3{\bf q}_1 \over (2\pi)^3}\int {d^2{\bf q}_2 \over (2\pi)^3} \nn
B_3(-{\bf k}_a,{\bf k}_a-{\bf q}_1, {\bf k}_b, {\bf k}_b-{\bf q}_2)\\
&& \hspace{4cm} \times W_L(\bq_1)W_L(\bq_2) \\
&& {\cal T}_{22}(\bk_1,\bk_2) =\la P({\bf k}_a)P({\bf k}_b,{\bf r}_L) \ra_c =4\,R_b \, P(k_a)\,P(k_b) \sigma^2_L \delta_D(k_{12}).
\label{eq:cross_power}
\een
Eq.(\ref{eq:cross_power}) is an estimate of the covariance of the local power spectrum.
We have used Eq.(\ref{eq:squeezed2}) in our derivation.

To define the integrated trispectra we can integrate the angular dependence of the vertices
$R_a$ and $R_b$ in the Fourier space in a way similar to the bispectrum case. 

The integrated trispectra ${\cal T}_{31}(\bk_a,\bk_b,\bk_c)$ and  ${\cal T}_{22}(\bk_a,\bk_b)$ defined in Eq.(\ref{eq:final_bispec})
and Eq.(\ref{eq:cross_power}) are related to the 
kurt-spectra i.e. $S_{31}$ and $S_{22}$ discussed previously.
 
\section{Integrated Bispectra: Quasilinar Regime}
\label{sec:unified_discussion}
In this section we will provide a unifying description of the integrated bispectrum
in various specific cases, 
e.g. the case of Exact Dynamics (ED), velocity divergence $\Theta$,
2D dynamics  and the Zel'dovich approximation (ZA).

\subsection{A Unifying Approach}
\label{subsec:unify}
In general a second-order effective kernel $X_2$ given below can describe both the 
density field $\delta$ and velocity divergence
$\Theta =\nabla\cdot {\bf v}/H$ statistics ($H$ is the Hubble parameter) for different choices of 
parameters of $\alpha$ and $\beta$:
\ben
&& X_2(\bk_1,\bk_2) = {\alpha}+{1 \over 2}({\alpha+\beta})\, 
\left( {k_1 \over k_2} + {k_2\over k_1}\right )
\left( {\bk_1\cdot\bk_2 \over k_1k_2} \right )
+ {\beta}\,\left( {\bk_1\cdot\bk_2 \over k_1k_2} \right )^2.
\label{eq:generic_bi}
\een
The above parametrization satisfies the constraint $X_2(\bk,-\bk)=0$ from momentum conservation (translational
invariance) \citep{Peebles}.
We have kept $\alpha$ and $\beta$ free but all physical models that we will consider satisfy $(\alpha+\beta)=1$.
\ben
&& X_2({\bf k}-{\bf q}_1,-{\bf q}_3) \approx
\alpha + {1 \over 2\,(kq_3)^2}(\alpha+\beta)\left [-({\bf k}\cdot {\bf q_3})k^2+({\bf q}_1\cdot {\bf q}_3)k^2 \right]
+ \beta \left({\bk\cdot\bq_3 \over kq_3} \right )^2  ; \nn \\
&&  X_2(-{\bf k}+{\bf q}_{13}, -{\bf q}_3)\approx {\alpha} + {1 \over 2 (kq_3)^2}(\alpha+\beta) 
\left [k^2 ({\bf k}\cdot {\bf q_3}) - 
k^2 ({\bf q}_1\cdot{\bf q}_{13}) \right ] + \beta\left ({{\bf k} \cdot {\bf q}_3 \over k q_3}\right )^2.\nn \\
\een
Imposing $\alpha+\beta=1$, the result for the squeezed bispectrum takes the following form: 
\ben
&& B_2(\bk-\bq_1,-\bk+\bq_{13},-\bk_3) \nn \\ 
&& \hspace{1cm}=\Big \{(3\alpha-\beta)+4\beta\left({\bk\cdot\bq_3\over kq_3}\right)^2
 - (\alpha+\beta)\left({\bk\cdot\bq_3\over kq_3}\right)^2{d\ln P(k)\over d\ln k} \Big \}P(k)P(q_3);
\label{eq:master}\\
&& = \left [ {3\alpha+{\beta \over 3} +1 }- {1\over 3}{d{\ln k^3P(k)} \over d\ln k }\right ]P(k)P(q_3)\quad ({\rm for\;\;3D}); \label{eq:master3D}\\
&& = \left [ {3\alpha+\beta +1 }- {1\over 2}{d{\ln k^2 P(k_\perp)} \over d\ln k_\perp }\right ]P(k_\perp)P(q_{\perp 3})\quad ({\rm for\;\;2D}).
\label{eq:master2D}
\een
The specific cases so far we have analysed in this paper are examples where the
triplets $\{ \alpha,\beta\}$ take the following values $\{5/7,2/7\}$ for ED
$\{1/2,,1/2\}$ for ZA and $\{3/7,4/7\}$ for  velocity divergence $\Theta$.
For projected density fields the angular averages need to be considered in 2D.
The actual bispectrum remains the same as 3D.

More complicated kernel where the parameters $\alpha,\beta$ are redshift $z$ and mode $k$
dependent provides better fit to numerical simulations and has
also been considered in the literature which can be incorporated in
this framework.

For a generic cosmology the kernels take the following form Ref.\cite{francis} (see Eq.(71) and Eq.(72) of this 
review Ref.\cite{francis} (Section:2.4.5); we have corrected a typo in Eq.(71)):
\ben
&& F_2(k_1,k_2) = {1\over 2}({1+\epsilon})+{1 \over 2}\, 
\left( {k_1 \over k_2} + {k_2\over k_1}\right )
\left( {\bk_1\cdot\bk_2 \over k_1k_2} \right )
+ {1\over 2}({1-\epsilon})\,\left( {\bk_1\cdot\bk_2 \over k_1k_2} \right )^2; \\
&& G_2(k_1,k_2) = \epsilon+{1 \over 2}\, 
\left( {k_1 \over k_2} + {k_2\over k_1}\right )
\left( {\bk_1\cdot\bk_2 \over k_1k_2} \right )
+ {(1-\epsilon)}\,\left( {\bk_1\cdot\bk_2 \over k_1 k_2} \right )^2.
\een
Here $\epsilon={3/7}\Omega_M^{-2/63}$ for $\Omega_{\rm M} \ge 0.1$ Ref.\cite{theta_skew}.
Using the generic expressions above in Eq.(\ref{eq:generic_bi}) we arrive at the following results for 3D:
\ben
&&\bar{\cal B}^{\delta}(k)= {1\over 3}\left [{(8 + 4\epsilon)} - (n+3)\right ]\sigma^2_LP_\delta(k);\\
&&\bar{\cal B}^{\theta}(k)= {1\over 3}\left [{(4 + 8\epsilon)} - (n+3)\right ]\sigma^2_LP_\delta(k).
\label{eq:theta3D}
\een
For $\Omega=1$ we recover $\bar{\cal B}^{\delta}(k) \equiv [{68/21}-(n+3)/3]$ and $\bar{\cal B}^{\Theta}(k) \equiv [{52/21}-(n+3)/3].$
For all practical purposed these results are sufficient as the dependence on $\Omega_M$ is is extremely weak.
For ZA we have $\bar{\cal B}^{\rm ZA,\delta}(k) \equiv [{8/3}-(n+3)/3]$ and $\bar{\cal B}^{\rm ZA,\Theta}(k) \equiv [{4/3}-(n+3)/3]$.
In case of $n=-3$ we recover the unsmoothed values  $\bar{\cal B}^{\delta}(k)=2\nu_2={68/21}$ and 
$\bar{\cal B}^{\Theta}(k)=2\mu_2={52/21}$. In comparison the skewness parameters are given by $S_3^\delta\equiv 3\nu_2$
and $S_3^{\Theta} = 3\mu_2$.
In 2D we have the following results:
\ben
&&\bar{\cal B}_{2D}^{\delta}(k) = \left [(2+\epsilon)-{1\over 2}(n+2)\right ]\sigma^2_LP_\delta(k_{\perp}); \\
&&\bar{\cal B}_{2D}^{\Theta}(k)= \left[ (2\epsilon+2)-{1\over 2}(n+2) \right ]\sigma^2_LP_\delta(k_{\perp}).
\een
For $\Omega=1$ we recover $\bar{\cal B}^{\delta}(k) = [(24/7)-(n+2)/2]\sigma^2_{\rm 2D,L}P_\delta(k_{\perp})$.

To linear order we have the well known result: $\Theta=-f(\Omega)\delta$. Using this in 
Eq.(\ref{eq:theta3D}) we obtain:
\ben
&&\bar{\cal B}^{\Theta}(k)= -{1\over 3\,f(\Omega)}\left [{(4 + 8\epsilon)} - (n+3)\right ]\sigma^2_{\Theta L}P_\Theta(k).
\label{eq:int_theta}
\een
Here, $f(\Omega) \approx \Omega^{3/5}$. This function is sensitive to any variation of $\Omega$
which makes the integrated bispectrum of $\Theta$ sensitive to $\Omega$, in contrast to $\delta$.
 
\subsection{Mixed $\delta-\Theta$ Integrated Bispectra}
\label{subsec:mixed}
In our analysis so far we have cross correlated the $\bar\delta$ and $P_\delta(k)$
as well as $\bar\Theta$ and $P_\Theta(k)$; these probe the squeezed {\em pure} bispectrum i.e.
$B_{\delta\delta\delta}$ or $B_{\Theta\Theta\Theta}$ but it is possible to device 
consistency tests by considering the mixed bispectra $B_{\delta\Theta\Theta}$ or
$B_{\Theta\delta\delta}$.  

Generalising Eq.(\ref{eq:final_bispec}) we introduce the following pair of mixed bispectra:
\ben
&& \la P_{\delta\delta}({k}) \bar\Theta({\bf r}_L)\ra_c= \quad {1 \over V_L^2}\int {d^3{\bf q}_1 \over (2\pi)^3}
\int {d^3{\bf q}_2\over (2\pi)^3}B_{\delta\delta\Theta}[{\bf k-q_1, -k+q_{12}, -q_2}]\nn \\
&& \hspace{3cm} \times W_L({\bf q}_1)W_L(-{\bf q}_{12})W_L(-{\bf q}_2);\\
&& \la P_{\Theta\Theta}({k}) \bar\Theta({\bf r}_L)\ra_c= \quad {1 \over V_L^2}\int {d^3{\bf q}_1 \over (2\pi)^3}
\int {d^3{\bf q}_2\over (2\pi)^3}B_{\Theta\Theta\delta}[{\bf k-q_1, -k+q_{12}, -q_2}]\nn \\
&& \hspace{3cm} \times W_L({\bf q}_1)W_L(-{\bf q}_{12})W_L(-{\bf q}_2).
\label{eq:mixed_bispec}
\een
\n
Going through the same algebra we can show:
\ben
&& \la P_{\delta\delta}({k}) \bar\Theta({\bf r}_L)\ra_c= f(\Omega)\left[{68\over 21}-{n+3\over 3} \right ]\sigma^2_L P(k);\\
&& \la P_{\Theta\Theta}({k}) \bar\delta({\bf r}_L)\ra_c= -f^2(\Omega)\left[{52\over 21}-{n+3\over 3} \right ]\sigma^2_L P(k).
\label{eq:mixed}
\een
Both expressions are sensitive to $\Omega$ owing to the presence of $\Theta$. Notice that the power spectra and 
the variance in these expressions are different compared to that in Eq.(\ref{eq:int_theta}).

Standard (Eulerian) Perturbation Theory (SPT) is known to agree well with numerical simulations for 
$z\ge 1$ and $k\le 0.2\rm h Mpc^{-1}$. They fail to provide accurate results in the highly non-linear regime
e.g. for the Baryon Acoustic Oscillation (BAOs) amplitudes at $k\ge 0.2 {\rm h} {\rm Mpc}^{-1}$. The SPT predictions
are redshift-independent, though in simulations BAOs show smaller amplitudes at lower redshift.
More accurate formula for the bispectrum exists \cite{Scoccimarro,GilMarin} which can be incorporated in 
our analysis.
Alternatively, the recently proposed separate Universe method can be employed
to compute the higher-order integrated spectra \citep{Baldauf,Creminelli,WayneHu,Pajer}.
In this approach the effect of long-wavelength density fluctuation on the small-scale power spectrum is computed
by treating each over- and under dense region as a separate universe
with a different background cosmology.

\subsection{Integrated Bispectra in Lagrangian Perturbation Theory}
\label{subsec:ZA}
The higher-order propagators take a particularly simpler form for the {\em Zeldovich Approimation} (ZA)
(see e.g. ref.\citep{MVS} and references therein).
The ZA is the first-order solution to perturbative dynamics formulated in Lagrangian space known
as the Lagrangian Perturbation Theory (LPT) \citep{review}. The second order kernel
that describes the ZA is given by the following expression:
\ben
F^{\rm ZA}_2({\bf q}_1,{\bf q}_2)={1\over 2} + {1\over 2}(\bq_1 \cdot \bq_2)\left ({1\over q_1^2}+{1\over q_2^2} \right ) +
{1\over 2}\left({\bq_1\cdot\bq_2 \over q_1\,q_2}\right)^2.
\een
This is a special case of the generic bispectrum studied in Eq.(\ref{eq:generic_bi}) 
for $\{\alpha,\beta \}=\{1/2,1/2\}$.
Using these expressions we can deduce the expression for the squeezed bispectrum in the leading
order as:
\ben
&& B_{\rm ZA}({\bf k}_{}-{\bf q}_1, -{\bf k}_{}+{\bf q}_{13}, -{\bf q}_{3})\nn \\
&& = \left [ 1 + 2 \left ( {{\bf k}\cdot {\bf q}_3 \over k q_3} \right )^2 -
\left ( {{\bf k}\cdot {\bf q}_3 \over k q_3} \right )^2 {d\ln P(k_{}) \over d\ln k_{}}  \right ] P(k)P(q_3)+
{\cal O}(q_3/k).
\een
The ZA and its higher-order analogues are often used to set-up the initial conditions in a numerical simulation.
The results can be derived using the same steps followed in the derivation of results from Eulerian perturbative dynamics
Eq.(\ref{eq:c21}) and Eq.(\ref{eq:c31}).
We quote the results here:
\ben
&& {\rm C}^{\rm ZA}_{21} = \left [ {8\over 3} -{1 \over 3}{d \ln k^3 P(k) \over d\ln k } \right ]P(k)P(q_3) = 
\left [{8 \over 3}-{(n+3)\over 3} \right ] P(k)P(q_3).
\label{eq:za}
\een
Eq.(\ref{eq:za}) is a special case of the general result presented in Eq.(\ref{eq:master}) for $\{\alpha,\beta\}=\{1/2,1/2\}$.
These can be used to gauge the level of transients arising from the initial conditions
often used in numerical simulations.
It is possible to compute the corrections from higher order LPT following
the same procedure (see e.g. \citep{MVS}). Squeezed configurations of the trispectrum
can also be computed in a similar manner. The higher order
kernels for the ZA are given in Eq.(\ref{eq:za_approx}).
\begin{figure}
\vspace{1.25cm}
\begin{center}
{\epsfxsize=13. cm \epsfysize=5. cm 
{\epsfbox[32 411 547 585]{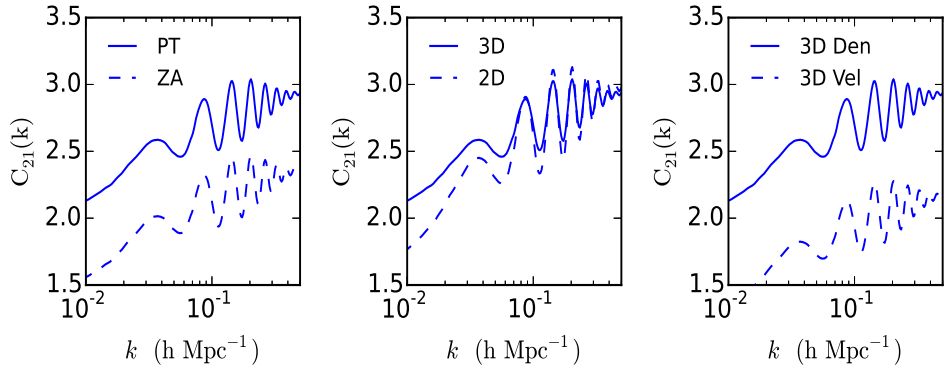}}}
\caption{
The left panel shows the integrated bispectrum from second-order Eulerian perturbation theory
and the lowest order Lagrangian perturbation theory, the ZA, following Eq.(\ref{eq:za}).
The middle panel compares the integrated bispectrum for 3D and 2D surveys Eq.(\ref{eq:2D}). Finally, the
right panel compares the integrated bispectrum for the density $\delta$ and the divergence of $\Theta$.}
\label{fig:cumu3}
\end{center}
\end{figure}

The integrated bispectrum for the ZA is presented in the left panel of Figure \ref{fig:cumu3}.
The solid curve shows the prediction from second order SPT and the dashed line represents the ZA.
For the entire range of $k$, the ZA under predicts the integrated bispectrum. This is related
to the fact that the vertex $\nu_2=4/3$ for ZA as compared to $\nu_2=34/21$ for the exact dynamics.
This values are consistent with skewness parameter $S_3 = 3\nu_2= 34/7$ for SPT and $S_3 = 4$ for ZA \cite{MVS}.
For $n=-3$ we recover the limit ${\rm C}_{21}=2\nu_2 = 8/3$. Finally, using Eq.(\ref{eq:define_ibispec}), the integrated bispectrum for
the ZA takes the following form:
\ben
\bar{\cal B}_{ZA}(k)= \left [{8 \over 3}-{(n+3)\over 3} \right ] P(k)\sigma^2_L.
\een   
\section{Integrated Bispectrum from Projected (2D) surveys}
\label{subsec:2D}
In this Section, we generalise the expression derived in 3D above to 2D or projected surveys. We consider
2D weak lensing surveys and 2D projected galaxy surveys. Though we eventually specialise the
results to projected galaxy surveys, the results are
equally relevant for studies of weak lensing and CMB secondaries (e.g. for the thermal Sunyaev Zeldovich (tSZ) effect).
The results derived here can also generalised to {\em cross-correlation} of two different surveys
or for {\em tomographic} analysis.

We start by defining an arbitrary projected field $\psi({\bm\gamma})$ defined on the surface of the sky 
obtained through the line-of-sight integration of the 3D field $\Psi(r,{\bm\gamma})$:
\ben
&& \psi({\bm\gamma}) =\int_0^{r_s} dr\, w(r)\,\Psi(r,\bm\gamma); \nn\\
&& \psi(\bm\gamma) = 
\int_0^{r_s}\, dr\, w(r)\, \int {d^3k \over (2\pi)^3} \exp[i(\,r\,k_{\parallel}+d_A(r)\bm\gamma\cdot \bk_{\perp})] 
\Psi(\bk).
\een
Here $r$ is the comoving radial distance and $d_A(r)$ is the comoving angular diameter distance. $\omega$ is
a generic radial selection function.
$k_{\parallel}$ and $\bk_{\perp}=d_{A}(r)\bell$ are the radial and projected components of the wave-vector $\bk$.

We will use small angle approximation (also known as the plane parallel approximation or the
distant observer approximation).
The average of a projected field $\psi({\bm\gamma})$ on the surface of the sky (${\bm\gamma}$ here represents unit vector
along a specific direction) is defined as:
\ben
\bar\psi({\bm{\gamma}}_0) = {1 \over {\Omega}}\int d^2{\bm\gamma}\, 
\psi({\bm\gamma})W_{\rm 2D}({{\bm\gamma}-{\bm\gamma}}_{0}); \quad {\Omega}= \int  d^2{\bm\gamma}.
\een
Here $W_{2D}$ is the 2D mask that encodes the sky coverage and $\Omega$ is the area of the sky covered. 
The window function defined as $W_{2D}({\bm \gamma}) \equiv \prod_{i=1}^{i=2}\theta({\bm\gamma}_i)$. The one-dimensional unit step
functions are the same as the ones defined in the 3D context in the previous section.
The 2D Fourier transform assuming a flat sky takes the following form:
\ben
\bar\psi({\bm\bell}, {\bm\gamma}_{0}) = 
\int d^2{\bm\bell^{\prime}}\psi({{\bm\bell}-{\bm\bell}^{\prime}})\, W_{\rm 2D}({\bell^{\bm\prime}})\, 
\exp(-i{\bm\gamma}_{0}\cdot {\bm\bell}).
\een
The 2D power spectrum in this fraction of sky is given by:
\ben
&& P_{\rm 2D}({\bm\ell},{{{\bm\gamma}}_0}) = {1 \over {\Omega}_{}} \int {d^2 {\bm\ell}_1 \over (2\pi)^2} 
\int {d^2 {\bm\ell}_2 \over (2\pi)^2}
\psi({\bm\ell}-{\bm\ell}_1) \psi(-{\bm\ell}-{\bm\ell}_2) \nn \\
&&\quad\quad\quad\quad\times \exp(-i\, {\bm\gamma}_0\cdot({\bm\ell}_1 + {\bm\ell}_2)) W_{\rm 2D}({\bm\ell}_1) W_{\rm 2D}({\bm\ell}_2).
\een
The resulting integrated bispectrum is defined by cross-correlating the local estimate
of the power spectrum and the local average of the projected field. 
\ben
&& {\cal B}_{\rm 2D}({\bell}) \equiv \la P_{\rm 2D}({\bell},{\bm\gamma}_{0})\bar\delta({\bm\gamma}_{0}) \ra; \\
&& {\cal B}_{\rm 2D}({\bell}) = {1 \over {\Omega}^2}  
\int {d^2{\bm\gamma} \over 4\pi}  \int {d^2 {\bell}_1 \over (2\pi)^2} \int {d^2 {\bell}_2 \over (2\pi)^2}
\la \psi({\bell}-{\bell}_1) \psi(-{\bell}-{\bell}_2)\psi(-{\bell}_3) \ra\nn \\
&& \hskip5cm \times\exp(-i\, {\bm\gamma}_0 \cdot ({\bell}_1+{\bell}_2+{\bell}_3)).
\een
The projected power spectrum $P_{\rm 2D}(\bell)$ and bispectrum $B_{\rm 2D}({\bell}_1,{\bell}_2,{\bell}_3)$ can be expressed in terms of the 3D power spectrum $P_{\rm 3D}(k)$ and bispectrum 
$B_{\rm 3D}({\bf k}_1,{\bf k}_2,{\bf k}_3)$: 
\ben
&& P_{\rm 2D}(\ell)\equiv \int_0^{r_s} d\,r {\omega^2(r) \over d^4_A(r)} 
P_{3D}\left ({\ell\over d_A(r)} \right );\\
&& B_{\rm 2D}(\bell_1,\bell_2,\bell_3) \equiv \int_0^{r_s} d\,r {\omega^3(r) \over d^6_A(r)}
B_{\rm 3D}\left ({\bell_1\over d_A(r)},{\bell_2\over d_A(r)},
{\bell_3 \over d_A(r)} \right)_{\sum {\bell}_i=0};
\een
see \citep{TakadaJain} and reference therein.
The expression for $B_{3D}$ is given in Eq.(\ref{eq:2Dbispec}). 
The angular average of the integrated bispectrum in 2D can be defined as follows:
\ben
\bar{\cal B}_{\rm 2D}(\ell) = \int {d\theta_\ell\over 2\pi}\; {\cal B}_{\rm 2D}(\bell); \quad \ell=|{\bell}|.
\een
The complete expression takes the following form:
\ben
B_{\rm 2D}(\ell) \equiv \int {d \theta_\ell \over 2\pi}\int {d^2{\bell}_{1} \over (2\pi)^2 }
\int {d^2{\ell}_{3} \over (2\pi)^2 }
B_{\rm 2D}(\bell-{\bell}_{1}, -{\bell}+{\bell}_{1}+{\bell}_{3}, -{\bell}_{3}).
\een
In the {\em squeezed limit} the bispectrum takes following form:
\ben
&& B_{\rm 2D}(\bell-{\bell}_{1}, -\bell+{\bell}_{1}+{\bell}_{3}, -{\bell}_{3})\nn \\
&& = \left [ {13 \over 7} + {8 \over 7}\left ( {\bell\cdot {\bell}_3 \over \ell \ell_3} \right )^2 -
\left ( {{\bell}\cdot {\bell}_3 \over \ell\ell_3} \right )^2 {d\ln P_{\rm 2D}(\ell) \over d\ln \bell}  \right ] 
P(\bell)P(\bell_3)+ \cdots
\een
The terms of higher order in $(\ell_1/\ell)$ or $(\ell_3/\ell)$ are ignored as we take the
limiting case when $\ell \gg\ell_i$.
Using the fact that the circular average of $\hat {\bell} \cdot {\hat {\bell}}_3$ is $[{1/2}]$
we arrive at the following expression:
\ben
&& \bar{\cal B}_{\rm 2D}(\ell) = K_3 \left [ {24 \over 7} - {1 \over 2} {d\ln \ell^2 P(\ell) \over d\ln \ell} \right ] 
P_{\rm 2D}(\ell)\sigma^2(\theta_0);\quad\quad\\
&&  K_3=\int_0^{r_s} dr {w^3(r)\over d_A^{(6+2n)}(r)}{\Big /}\left [\int_0^{r_s} dr {w^2(r)\over d_A^{(4+n)}(r)}\right ]^2.
\label{eq:2D}
\een
This matches the published results on cumulant correlators quoted below in Eq.(\ref{eq:2Dcumu})
for a 3D power spectrum which can be described locally as a power-law with a slope $n$ i.e. $P(k)\propto k^n$.
The corresponding cumulant correlators are derived in \cite{2Dcumu}:
\ben
\label{eq:2Dcumu}
&& {\rm C}^{\rm 2D}_{21}={24 \over 7} - {1 \over 2}(n+2); \\
&& {\rm C}^{\rm 2D}_{31}= {1473 \over 49} - {195 \over 14}(n+2) + {3 \over 2}(n+2)^2.
\een
Using very similar arguments we can show that if we assume a HA for the
underlying 3D bispectrum Eq.(\ref{eq:b2_all_k}), the corresponding integrated bispectrum is given by:
\ben
\bar{\cal B}_{\rm 2D}(\ell) = 2K_3 \,Q_3\, P_{\rm2D}(\ell)\sigma_{\rm 2D}^2(\theta_0); \quad \sigma_{\rm 2D}^2(\theta_0)\equiv\int{d\ell \over 4\pi}\ell P(\ell)W_{\rm 2D}^2(\ell\theta_0).
\label{eq:multi}
\een
The integrated bispectrum $\bar{\cal B}_{\rm 2D}(\ell)$
in 2D is plotted in Fig.\ref{fig:cumu2} as a function of $\ell$ (middle panel). The expression for the multiplicative
factor $K_3$ in Eq.(\ref{eq:multi}) depends
on the survey geometry and selection function which is not included in the plot.

These results can readily be extended to the case of two different surveys with overlapping sky coverage 
but different radial selection functions or for surveys with tomographic bins.

\section{Results and Discussion} 
\label{sec:res_disc}
The position-dependent power spectrum,
a probe of squeezed configuration of bispectrum,
was recently proposed as a method to probe galaxy clustering.
Cumulant correlators and their Fourier transform, the skew-spectra, are
also often used to probe the primary or secondary non-Gaussianity. 
In this paper, we have compared these two techniques and elucidated their
relationship to one another.
 
First, we have generalised the
concept of skew-spectrum and kurt-spectrum defined at third and fourth-order
to arbitrary order. We used known perturbative results to
show [Eq.(\ref{eq:most_important_result})] in the large separation limit,
or low $\bk$ limit ($\bk\rightarrow 0$), the generalisations of 
skew-spectra defined in Eq.(\ref{eq:s21}) to higher-order, also known as the multispectra $S_{pq}(k)$, 
are proportional to the underlying power spectrum with proportionality constants 
${\rm C}_{pq}$ [see e.g. Eq.(\ref{eq:c21})]
that are known to arbitrary order. The proportionality constants depend
on the local (linear) power-spectral index $n$ at the smoothing scale and can be 
computed to arbitrary order. These coefficients, deduced using a top-hat smoothing window, 
are known in 2D and 3D,
and are related to two-point joint PDFs $p_{\delta}(\delta_1,\delta_2)$ or equivalently the bias $b(\delta)$,
defined in Eq.(\ref{eq:bias}), of overdense objects. The computation of $S_{pq}(k)$ for the entire range of $k$ 
requires numerical evaluation. This has been carried out in for the $S_{21}(k)$ in Ref.\cite{PrattenMunshi}
for 3D galaxy surveys. Notice that the skew-spectra and kurt-spectra have also been
employed in analysing primordial non-Gaussianity in CMB temperature 
maps (ref.\cite{skewspec,kurtspec}). However the mulrispectra that we consider here are sub-optimal, 
where as, for CMB studies {\em optimised} versions were considered to improve their sensitivity
to primordial non-Gaussianity. 

Next, we generalised the 
concept of a position-dependent power spectrum or integrated bispectrum (IB) of the density field $\delta$
in many directions. 
We use a unifying approach in \textsection{\ref{sec:unified_discussion}} to investigate IB.
Using a generic bispectrum Eq.(\ref{eq:generic_bi})
we have deduced the IB for $\delta$ and $\Theta$ in Eq.(\ref{eq:theta3D})
from a master Eq.(\ref{eq:master}) that can also deal, with the 
bispectrum from lowest order of Lagrangian perturbation theory, the ZA.
Using Limber's approximation, we have also applied this result to projected (2D) surveys 
in \textsection{\ref{subsec:2D}}. These results can be readily generalised to
tomographic surveys or to cross-correlation of overlapping surveys using two different tracer fields.
Extending the concept of IB for one field 
we have generalised it to consider ($\delta$-$\Theta$) mixed bispectrum in \textsection{\ref{subsec:mixed}}.
In Eq.(\ref{eq:int_theta})-Eq.(\ref{eq:mixed}) we have pointed out that such measurements are sensitive
to cosmological
parameter $\Omega$. The results 
for ZA will particularly be useful in assessing magnitude of transients in numerical simulation.
Using the unifying approach, we were able to show that 
in each of these specific cases the expressions for ${\rm C}_{21}(k)$ and $\rm R_{21}(k)$ share
the same analytical expression Eq.(\ref{eq:master}).  
Despite the formal mathematical
similarities, the actual interpretation is quite different. In case of cumulant correlator ${\rm C}_{21}$
a given smoothing scale dictates the spectral index $n$. To map out the entire range of $k$
a range of smoothing scales are needed. Similarly, the momentum-dependence of the integrated bispectrum 
can be studied using many sub-samples of the survey and taking an approximate ensemble average.
Both methods can be used simultaneously as a consistency check. The power law $n=-3$ 
correspond to the case of no smoothing. In this case we recover the scale independent HA value of
$2\bar Q \equiv 2\nu_2 = {68/21}$ using the angular average of $Q$ i.e. $\bar Q = {34/ 21}$ [see eq.(\ref{eq:q3})].
Notice that this is true also for 2D and divergence of velocity $\Theta$.
In the case of $\Theta$, the unsmoothed vertex takes the numerical value: $Q_3=2\bar G = 52/21$. 

Going beyond second-order in Standard (Eulerian) Perturbation Theory (SPT) we have extended the
concept of IB to integrated trispectrum (IT) in Appendix-\textsection\ref{sec:tri}. We introduced two
ITs at the level of trispectrum: ${\cal B}_{22}(k)$ and ${\cal B}_{31}(k)$ 
respectively in Appendix-\textsection{\ref{sec:tri_collapsed}} and Appendix\textsection{\ref{sec:tri_squeezed}}.
In the {\em soft} limit they correspond to squeezed and collapsed
limits of the trispectrum. They are analogues of the corresponding
cumulant correlators ${\rm C}_{22}={\rm C}^2_{21}$ and ${\rm C}_{31}$ respectively.
The IT ${\cal B}_{22}(k)$ can be constructed using the expression for the ${\cal B}_{21}(k)$ 
[Eq.(\ref{eq:B21})]
and shows a structural similarity with ${\rm C}_{22}$. 
The explicit evaluation of ${\cal B}_{31}(k)$ was recently performed in Ref.\cite{komatsu2}.
However, the functional form for ${\cal B}_{31}(k)$ [Eq.(\ref{eq:R3})-Eq.(\ref{eq:R3E})] is not same
as that of ${\rm C}_{31}(k)$. We expect the same to be 
true for higher order integrated spectra. However, extension
of these results to higher orders can be cumbersome
owing to the complicated structure of the higher-order kernels $F_n$ see Eq.(\ref{eq:kerF}).
We conclude that higher-order multispectra and higher-order integrated spectra
can provide complementary information and much needed consistency checks
on probes of non-Gaussianity in diverse cosmological data sets. 
 
In addition to the SPT and LPT we have used the HA to get insight
into soft limits of higher order polyspectra [Eq.(\ref{eq:two})-Eq.(\ref{eq:five})]. In HA the tree perturbative
hierarchy is replaced with a similar hierarchy, but where the
kernels $F_n$ and $G_n$ 
are replaced by vertices $\nu_n$ and $\mu_n$ which are
angular averages of these kernels [Eq.(\ref{eq:nu})-Eq.(\ref{eq:mu})]. 
Many different models of HA exist and it is indeed possible also to leave these vertices as unknown parameters.
This model is only valid 
in the highly non-linear regime and thus strictly speaking not suitable for 
taking ${\bf k} \rightarrow 0$. However, it provides very useful
insight in higher order where exact SPT results are prohibitively
complicated especially in an idealised situation of $n=-3$ when
smoothing can be ignored. The squeezed limit for the HA bispectrum
is given in Eq.(\ref{eq:b2_first_k}) and the collapsed and squeezed limits
of the trispectrum are presented in Eq.(\ref{eq:b3_first_k}) and Eq.(\ref{eq:b3_second_k}) respectively.

The CCs and higher order integrated spectra both depend only on one wave number so
they are much easier to estimate than the corresponding full polyspectra. 
It is also much simpler to compute their covariance.

In this paper we have primarily focused on the theoretical aspects 
of IB and IT. We have shown that with other related statistics CCs and integrated spectra
can play complementary rule in probing soft limits of higher order polyspectra in 3D or projection (2D). 

However, to use the estimators proposed here it will be important to
develop them further. For example it's important to include redshift space distortion
to analyse galaxy surveys - which will involve analysing soft limits of
polyspectra in redshift space \citep{galaxy_red}.
Our results here are based on perturbative analysis, but, including results from halo model
can be done in a relatively straightforward manner to extend the
range of validity. Similarly, it is not difficult to extend the
results here to include primordial non-Gaussianity, though they remain
highly constrained by recent CMB observations \cite{Planck} at least at scales
probed by CMB observations. 
 
For weak lensing surveys, going beyond the 2D or tomographic analysis presented here it is 
now becoming practical to analyse the
data in 3D. Weak lensing probes structure formation at small scale. Gravity induced non-Gaussianity
is known to provide additional information to constrain the cosmology.
Our approach developed here can be
generalised to 3D weak lensing surveys using a spherical Fourier-Bessel transformation
\cite{3Dlens1,3Dlens2}.
In the field of CMB research, squeezed configuration of primordial non-Gaussianity
and its effect on CMB lensing have been investigated \cite{sq_cmb,sq_cmb1}. 
Two important secondaries - the lensing of CMB \cite{CMB_lensing}, and the
kSZ effect - both have a vanishing bispectrum \cite{kSZ}. They do not have any frequency 
information either. Thus the two sets of IT discussed here
can be useful in separating these two secondaries.
Results presented here will also be useful in analysing frequency-cleaned 
$y$-parameter maps \citep{tSZ1} or to study squeezed limit of
bispectrum induced by reionization \citep{Pandolfi}.
These estimators can also generalised to cross-correlate weak-lensing $\kappa$ maps
and $y$ maps \cite{tSZ2}. The {\em separate universe} approach developed 
by several authors remain a possibility for such development \citep{Baldauf,Creminelli,WayneHu,Pajer}.
Indeed the morphological estimators or the Minkowski Functionals (MF)
are a popular method to study non-Gaussianity in cosmological fields.
MFs depend on the higher order polyspectra and squeezed limit of
polyspectra can also be related to the position dependent MFs. 
The study of soft limit of polyspectra for CMB secondaries may provide a method to test the
kinematic consistency relations to constrain modified gravity theories
or primordial non-Gaussianity \cite{valag_consistency}.   

Estimation of integrated spectra (IB or IT)
is undoubtedly simpler than the corresponding polyspectra,
but designing optimal estimators to extract information about 
higher-order non-Gaussianities it is not a simple task. A particular 
difficulty is posed by the need to estimate the sample variance arising from
the survey. The scatter in the IB we deduced in this paper 
used a very simple prescription that ignores the very non-Gaussianity we
seek to characterise. In a regime in which the approximation of 
mild non-Gaussianity breaks down such a treatment will become
inadequate.

Finally, note that the estimators developed here are sub-optimal.
Though may not be too serious a concern for high quality data sets 
but in any case they are valuable by virtue of being much easier
to implement in practice than optimal estimators. 

\section{Acknowledgements}
\label{acknow}
DM and PC acknowledge support from the Science and Technology
Facilities Council (grant number ST/L000652/1).

\appendix
\section{Beyond the Integrated Bispectrum (IB): the Integrated Trispectrum (IT)}
\label{sec:tri}
We will briefly quote some results from Standard (Eulerian) Perturbation Theory (SPT)
that are relevant in our context.
The perturbative expansions of the density field $\delta$ and $\Theta$ can be 
expressed in terms of kernels  $F_n$ and $G_n$: 
\ben
&& \delta(\bk) = \delta^{(1)}(\bk)+\delta^{(2)}(\bk)+ \cdots; \nn \\
&& \delta^{(n)}(\bk) = \int d^3 \bq_1 \cdots \int  d^3 \bq_n F_n(\bq_1,\cdots, \bq_n) \delta(\bq_1)\cdots \delta(\bq_n).\\
&& \Theta(\bk) = \Theta^{(1)}(\bk)+\Theta^{(2)}(\bk)+ \cdots; \nn \\
&& \Theta^{(n)}(\bk) = \int d^3 \bq_1 \cdots \int  d^3 \bq_n G_n(\bq_1,\cdots, \bq_n) \delta(\bq_1)\cdots \delta(\bq_n).
\een
The expressions for the $n$th order kernels
$F_n$ and $G_n$ for $\delta$ and $\Theta$ respectively are Ref.\citep{review}:
\ben
&& F_n({\bf q}_1, \cdots, {\bf q}_n) =
\sum_{m=1}^{n-1}{G_m({\bf q}_1,\cdots, {\bf q}_m) \over (2n+3)(n-1)}
[(2n+1)\alpha({\bf k}_1,{\bf k}_2) F_{n-m}({\bf q}_{m+1},\cdots,{\bf q}_{n})\nn \\
&& \hspace{4cm} + 2\beta({\bf k}_1,{\bf k}_2)G_{n-m}({\bf q}_{m+1}, \cdots, {\bf q}_n)].
\label{eq:kerF}
\een
\ben
&& G_n({\bf q}_1, \cdots, {\bf q}_n) =
\sum_{m=1}^{n-1}{G_m({\bf q}_1,\cdots, {\bf q}_m) \over (2n+3)(n-1)}
[3\alpha({\bf k}_1,{\bf k}_2) F_{n-m}({\bf q}_{m+1},\cdots,{\bf q}_{n}) \nn \\
&& \hspace{4cm} + 
2n\beta({\bf k}_1,{\bf k}_2)G_{n-m}({\bf q}_{m+1}, \cdots, {\bf q}_n)].
\label{eq:kerG}
\een
Here $F_1=1$ and $G_1=1$ and the functions $\alpha$ and $\beta$ are defined as:
\ben
\alpha({\bf k}_1,{\bf k}_2) \equiv {{\bf k}_{12} \cdot {\bf k}_1 \over k_1^2};
\;\;\;
\beta({\bf k}_1,{\bf k_2}) \equiv k^2_{12} {{\bf k}_1 \cdot {\bf k}_2 \over 2k_1^2 k_2^2}.
\label{eq:alphabeta}
\een
We have defined the following quantities above:
\ben
{\bf k}_1 = {\bf q}_1+ \cdots + {\bf q}_m; \;\;\; 
{\bf k}_2 = {\bf q}_{m+1} + \cdots + {\bf q}_n; \;\;\;
{\bf k}= {\bf k}_1 + {\bf k}_2.
\een
The vertices $F_n$ for the lowest order Lagrangian Perturbation Theory (LPT) or
ZA take the following form:
\ben
&& F_n({\bf q}_1,\cdots,{\bf q}_n) = {1 \over n!} {{\bf k}\cdot {\bf q}_1 \over q_1^2}\cdots {{\bf k}\cdot {\bf q}_n \over q_n^2};\quad\quad {\bf k} \equiv {{\bf q}_1+\cdots + {\bf q}_n}.
\label{eq:za_approx}
\een
The angular averages of the kernels are the tree-levels amplitudes or the vertices as defined below:
\ben
\label{eq:nu}
&& \nu_n \equiv n! \int {d\oh_1 \over 4\pi}\cdots \int {d\oh_n \over 4\pi}
F_n(\bk_1,\cdots \bk_n); \\
&& \mu_n \equiv n! \int {d\oh_1 \over 4\pi}\cdots \int {d\oh_n \over 4\pi}
G_n(\bk_1,\cdots \bk_n).
\label{eq:mu}
\een
Using Eq.(\ref{eq:kerF}) and Eq.(\ref{eq:kerG}) the second and third order kernels are defined as follows:
\ben
&& F_2({\bf k}_1,{\bf k}_2) \equiv {5 \over 7} + {1 \over 2}\left ( {1 \over k_1^2} + {1 \over k_2^2} \right )({\bf k}_1\cdot{\bf k}_2)
+ {2 \over 7}{({\bf k}_1 \cdot {\bf k}_2 )^2 \over k_1^2 k_2^2}; \\
&& F_3({\bf k}_1,{\bf k}_2, {\bf k}_3) = {7 \over 18} {{\bf k}_{12} \cdot {\bf k}_1 \over k_1^2} \left [ F_2({\bf k}_2,{\bf k}_3)+G_2({\bf k}_1,{\bf k}_2) \right ]  \nn\\
&& \hspace{2.4cm} +{2 \over 18} {{\bf k}_{12}^2 ({\bf k}_1 \cdot {\bf k}_2) \over k_1^2 k_2^2}\left [ G_2({\bf k}_2,{\bf k}_3)
+ G_2({\bf k}_1,{\bf k}_2) \right ].\\
&& G_2({\bf k}_1,{\bf k}_2) \equiv {3 \over 7} + {1 \over 2}\left ( {1 \over k_1^2} + {1 \over k_2^2} \right )({\bf k}_1\cdot{\bf k}_2)
+ {4 \over 7}{({\bf k}_1 \cdot {\bf k}_2 )^2 \over k_1^2 k_2^2}.
\een
\n
Using the fact that in 3D the angular averages of $\alpha$ and $\beta$ 
are respectively $\bar \alpha =1$ and $\bar\beta={1 \over 3}$ we obtain:
\ben
&& \nu_2\equiv 2\bar F_2 = 2\left [{5 \over 7}+{2 \over 7}{1 \over 3}\right ]= {34 \over 21}; \quad\quad
\mu_2 \equiv 2\bar G_2 = 2\left[ {3 \over 7}+{4 \over 7}{1 \over 3}\right ]= {26 \over 21};\quad \\
&& \nu_3\equiv 6\bar F_3 = 6\left[ {7\over 18}\left ( {17\over 21}+{13 \over 21}\right ) + {4 \over 18}\cdot{1 \over 3}\cdot{13 \over 21} \right] = {682 \over 189}.
\label{eq:vertices}
\een
For 2D we use $\bar \alpha =1$ and $\bar\beta={1 \over 2}$;
in this case we have  $\nu_2 \equiv 2{\bar F_2}= {12\over 7}$, $\mu_2 \equiv 2{\bar G_2}= {10\over 7}$,

Following recursion relation can be derived using Eq.(\ref{eq:kerF}) and Eq.(\ref{eq:kerG}) that is useful in
evaluation of $\nu_n$ and $\mu_n$ results quoted above:
\ben
\nu_n = \sum_{m=1}^{n-1}{n\choose m} {\mu_m \over (2n+3)(n-1)} 
\left [ (2n+1)\nu_{n-m} + {2 \over 3}\mu_{n-m} \right ];\\
\mu_n = \sum_{m=1}^{n-1}{n\choose m} {\mu_m \over (2n+3)(n-1)}
\left [ 3\nu_{n-m} + {2 \over 3}n\mu_{n-m} \right ].
\een
The perturbative bispectrum $B^{\rm PT}(\bk_1,\bk_2,\bk_3)$ and 
trispectrum  $T^{\rm PT}(\bk_1,\bk_2,\bk_3,\bk_4)$ take the following forms:
\ben
&& B^{\rm PT}({\bf k}_1,{\bf k}_2,{\bf k}_3) = 2F_2({\bf k}_1,{\bf k}_2)P(k_1)P(k_2) + 2\; {\rm perm.}; \label{eq:2Dbispec}\\
&& T^{\rm PT}({\bf k}_1,{\bf k}_2,{\bf k}_3,{\bf k}_4) =  
4 \left [ F_2({\bf k}_{13},-{\bf k}_1) F_2({\bf k}_{13},{\bf k}_2)  P(k_{13})P(k_2)P(k_2) + 11\; {\rm perm.} \right ] \nn \\
&& \quad\quad\quad\quad + 6 \left [ F_3({\bf k}_1,{\bf k}_2,{\bf k}_3) P(k_1)P(k_2)P(k_3) + 3 {\rm perm.} \right ]
\een 
\section{Perturbative Computation of the {\em Collapsed} Trispectrum}
\label{sec:tri_collapsed}
The aim in this section is to deduce the normalisation coefficient for the collapsed trispectrum
and show it is same as given in Eq.(\ref{eq:c21}).
In the collapsed configuration the trispectrum includes contributions only from 
{\em snake} diagrams.
\ben
&& B_3(\bk_1,\bk_2,\bk_3,\bk_4) =
\la\delta(\bk_1)\delta^{(2)}(\bk_2)\delta^{(2)}(\bk_3)\delta(\bk_4)\ra_c
+ \la\delta(\bk_2)\delta^{(1)}(\bk_2)\delta^{(2)}(\bk_3)\delta(\bk_4)\ra_c \nn\\
&& \hspace{2cm}+\la\delta(\bk_1)\delta^{(2)}(\bk_2)\delta^{(2)}(\bk_4)\delta(\bk_3)\ra_c
+ \la\delta(\bk_2)\delta^{(1)}(\bk_2)\delta^{(2)}(\bk_4)\delta(\bk_3)\ra_c.
\label{eq:bi}
\een
Following  Eq.(\ref{eq:kerF}) we express the second-order correction $\delta^{(2)}(\bk)$:
\ben
\delta^{(2)}(\bk) = \delta_{\rm 3D}(\bk-\bk_{ab})\int\int  {\rm F}_2(\bk_a,\bk_b)\delta^{(1)}(\bk_a)\delta^{(1)}(\bk_b) d^3\bk_a \, d^3\bk_b ;\quad \bk_{ab} = \bk_a+\bk_b
\een
Here $\delta_{3\rm D}$ is the 3D Dirac delta-function. Taking an ensemble average leads us to the following expression:
\ben
\la\delta(\bk_1)\delta^{(2)}(\bk_2)\delta^{(2)}(\bk_3)\delta(\bk_4)\ra_c = 
F_2(-\bk_2,\bk_{12})F_2(-\bk_4,-\bk_{12})P(\bk_1)P(\bk_{12})P(\bk_4).
\een
Combining the contributions from all four terms in Eq.(\ref{eq:bi}):
\ben
&& B_3(\bk_1,\bk_2,\bk_3,\bk_4) =
\rP(\bk_{12}) \left [ F_2(-\bk_1,\bk_{12}) \rP(\bk_1) + F_2(-\bk_2,\bk_{12}) {\rP}(\bk_2) \right ]\nn \\
&& \hspace{3cm}\times \left [ F_2(-\bk_3,\bk_{34}) \rP(\bk_3) + F_2(-\bk_4,\bk_{34}) \rP(\bk_4)\right ].
\een
We derive the expression for the collapsed trispectrum in this section.
The results will be of practical use in estimation of covariance of 
local power spectrum estimates from survey sub-volumes.
We start with definition of the local power spectrum in a sub-volume in Eq.(\ref{eq:local_ps}).
Next, we compute the covariance between the power spectrum at different 
mode $k$ and $k^{\prime}$:
\ben
&& \la \hat \rP({\bf k},{\bf r}_\rL)\hat \rP({\bf k}^{\prime},{\bf r}_\rL) \ra_c
= {1 \over V^2_L}\int {d^3 {\bf q}_1\over (2\pi)^3}
\int {d^3 {\bf q}_2\over (2\pi)^3}\int {d^3 {\bf q}^{\prime}_1\over (2\pi)^3}
\int {d^3 {\bf q}^{\prime}_2\over (2\pi)^3}\nn \\
&& \hspace{1cm}\times \la\delta(\bk-\bq_1)\delta(-\bk-\bq_2)
\delta(\bk^{\prime}-\bq^{\prime}_1)\delta(-\bk^{\prime}-\bq^{\prime}_2)\ra \nn \\
&& \hspace{1cm}\times W_{\rL}(\bq_1)W_{\rL}(\bq_2)W_{\rL}(\bq_1^{\prime})W_{\rL}(\bq_2^{\prime})
\exp[-i{\bf r}_{\rL}\cdot(\bq_{12}+\bq^{\prime}_{12})].
\een
We use the following definition of collapsed trispectrum:
\ben
&& \la\delta(\bk-\bq_1)\delta(-\bk-\bq_2)
\delta(\bk-\bq_1)\delta(-\bk-\bq)\ra_c \nn \\
&& =(2\pi)^3\delta_{\rm D}(\bq_{12}+\bq^{\prime}_{12})
{B}_3[{\bf k}-{\bf q}_1, -{\bf k}+{\bf q}_1+{\bf q}_2, 
{\bf k}^{\prime}-{\bf q}_1^{\prime},-{\bf k}^{\prime}+{\bf q}^{\prime}_1+{\bf q}^{\prime}_2]. 
\label{eq:tri1}
\een
In the collapsed limit the trispectrum takes the following form:
\ben
\lim_{\bq_i\rightarrow 0} {B}_3[{\bf k}-{\bf q}_1, -{\bf k}+{\bf q}_1+{\bf q}_2, 
{\bf k}^{\prime}-{\bf q}_1^{\prime},-{\bf k}^{\prime}+{\bf q}^{\prime}+{\bf q}^{\prime}_3] 
\stackrel{\text{collapsed}}{\approx} {B}_3[\bk, -\bk, \bk^{\prime}, -\bk^{\prime}].
\een
To simplify further, we express the 3D delta function $\delta_{3\rD}$ in Eq.(\ref{eq:tri1}) as a convolution
of two 3D delta function:
\ben
\delta_{3\rD}(\bq_{12}+\bq^{\prime}_{12}) = \int d^3 \bq_3\, 
\delta_{3\rD}(\bq_{12}+\bq_3)\,
\delta_{3\rD}(\bq^{\prime}_{12}-\bq_3).
\een
We use these $\delta_{3\rD}$ functions to collapse the $\bq_2$ and $\bq^{\prime}_2$ integrals:
\ben
&& \la \hat \rP(\bk,\br_\rL)\hat \rP(\bk^{\prime},\br_\rL)\ra_c={1 \over V_\rL^2} 
\int{d^3 \bq_1 \over (2\pi)^3} \int {d^3 \bq^{\prime}_1\over(2\pi)^3}
\int {d^3 \bq_3\over(2\pi)^3}\nn \\
&& \times {B}_3[\bk-\bq_1,-\bk-\bq_1-\bq_3, 
\bk^{\prime},-\bk^{\prime}-\bq^{\prime}_1+\bq_3]\nn  \\
&& \times W_L(\bq_1)W_L(-\bq_1-\bq_3)W_L(\bq_1^{\prime})W_L(\bq_1-\bq).
\een
After tedious but straightforward simplification, we get:
\ben
&& {B}_3^{\rm coll}(\bk_1,\bk_2) \equiv  {B}_3[{\bf k}-{\bf q}_1, 
-{\bf k}+{\bf q}_1+{\bf q}_3, 
{\bf k}^{\prime}-{\bf q}_1^{\prime},-{\bf k}^{\prime}+{\bf q}_1^{\prime}-{\bf q}_3] \nn \\
&& \vspace{2cm }=  {P}(k){P}(k^{\prime}){P}(q_3)
\left [{13 \over 7}+ 
{8 \over 7}\left({{\bf k}\cdot{\bf q}_3 \over k\, q_3}\right)^2
-\left({{\bf k}\cdot {\bf q}_3 \over k\, q_3}\right)^2 {d\ln { P}(k)\over d\ln k}\right ]
[{\bf k}\rightarrow {\bf k}^{\prime}].
\een
The expression in the second bracket is obtained by replacing ${\bf k}$ with ${\bf k}^{\prime}$.
Next, we perform the angular integrals in the Fourier space.
\ben
&&{B}_{3}^{\rm coll}(k,k^{\prime}) \equiv
\int {d^2\hat\Omega_{\bf k}\over 4\pi}\int {d^2\hat\Omega_{{\bf k}^{\prime}}\over 4\pi}
\,{B}^{\rm coll}_{3}(\bk,\bk^{\prime})\nn \\
&& = {P}(k){P}(k^{\prime})\sigma^2_L\left [{68 \over 21} - {1 \over 3} {d\ln k^3 P(k) \over d\ln k}\right ]
[{k} \rightarrow {k}^{\prime}].
\een
In our derivation, we have taken advantage of the Eq.(\ref{eq:window}). The factorisation
of the expression in terms of products of two factors that depend either on ${\bk}$ or ${\bk}^{\prime}$
allows us to perform the respective angular integration independently. 
Finally, assuming a local power-law for the power spectrum $P(k)\propto k^n$, we get:
\ben
&& {B}_3^{\rm coll}(k,k^{\prime})
\equiv  {P}(k){P}(k^{\prime})\,\sigma_L^2\,
\left [2\nu_2 - {1 \over 3} (n+3)\right ]
\left [2\nu_2 - {1 \over 3} (n^{\prime}+3)\right ]; \nn \\
&& {d\ln k^3 {P}(k) \over d\ln k} = (n+3); \quad\quad
\sigma_\rL^2 = {1 \over V_\rL^2} \int d^3\bq\, {\rm P}(k)\, W^2_\rL(\bq).
\label{eq:B21}
\een
The amplitude $\nu_2={34/21}$ is defined in Eq.(\ref{eq:vertices}). As expected this numerical coefficient is
identical to what was quoted for cumulant correlator in Eq.(\ref{eq:c21}). 
The factorization ${\rm C}_{22}={\rm C}^2_{21}$ is a result of tree-level perturbation theory.
Higher order contributions will be ${\cal O}(\sigma_L^4)$. For a reasonable 
big sub-volume such contribution will be negligible.

To recover the results derived in for HA valid in the non-linear regime
\textsection(\ref{sec:tri}) we have to set $n=-3$ and identify $R_a= \nu_2^2$ 
eq.(\ref{eq:cross_power}).
The results derived here assumes a $\Omega=1$ EdS cosmology. 
To probe residual dependence on cosmology we can follow the procedure outlined in \textsection\ref{subsec:unify}.
A similar derivations using $X_2(\bk_1,\bk_2)$ defined in Eq.(\ref{eq:generic_bi}) can be carried out 
which will replace the square bracket in Eq.(\ref{eq:B21})
with appropriate $\Omega$ dependence of Eq.(\ref{eq:master3D}) or Eq.(\ref{eq:master2D}) (in case of 2D). 
This will also generalise the above result also to the case of $\Theta$ or for the case of ZA. 
\section{Perturbative Computation of {\em Squeezed} Trispectrum}
\label{sec:tri_squeezed}
The aim of this section is to show that in the squeezed limit the normalisation coefficient takes same the form
as Eq.(\ref{eq:c31}). However in the squeezed configuration both star and snake diagrams contribute, thus making
the calculations more involved.

\noindent
{\bf Contributions From Snake Diagrams:}
The following six snake terms of the total twelve terms contribute in the leading order in the squeezed configuration: 
\ben
\label{eq:snake}
&&\lim_{\bq\rightarrow 0} {B}_3(\bq,\bk_2,\bk_3,\bk_4) \stackrel{\text{snake}}{=} \lim_{\bq \rightarrow 0}\rP(\bq)\Big \{ P(\bk_2)P(\bk_4) \rF_2(-\bk_2,-\bk_4)\left [ \rF_2(-\bq, \bk_2) + \bk_2\rightarrow \bk_4 \right ]\nn\\
&& \hspace{3cm} + \rP(\bk_3)\rP(\bk_4) \rF_2(-\bk_3,-\bk_4)\left [ \rF_2(-\bq, \bk_3) + \bk_3 \rightarrow \bk_4 \right ] \Big \} \nn \\
&& \hspace{3cm} + \rP(\bk_2)\rP(\bk_3) \rF_2(-\bk_3,-\bk_2)\left [ \rF_2(-\bq, \bk_2) + \bk_2\rightarrow \bk_3 \right ] \Big \}\delta_{3\rm D}(\bk_{234}).
\een
\ben
&& {B}_2(\bk_2,\bk_3,\bk_4) = {\rm F}^{\rm sq}_2(\bk_1,\bk_2)\rP(k_2)\rP(k_3) + \rm cyc. perm.;\\
&& {\rm F}^{\rm sq}({\bk_2,\bk_4})= {\rm F}_2(\bk_2,\bk_4) \left [ \rF_2(-\bq, \bk_2) + \bk_2 \rightarrow \bk_4 \right ]
\een
Thus the configuration from snake diagrams in the squeezed trispectrum
takes the form of a bispectrum with a different vertex amplitude ${\rm F}^{\rm sq}_2$.
For the hierarchical model the vertices are constant
 $F_2(\bk_1,\bk_2)=\nu_2$. In this limit the squeezed trispectrum takes simpler form
and can be expressed in terms of the hierarchical bispectrum:
\ben
\lim_{\bq\rightarrow 0}{B}_3(\bq,\bk_2,\bk_3,\bk_4) = 2\nu_2\, \rP(q)\,{B}_2(\bk_2,\bk_3,\bk_4)
\een
In the limit $\{\bq, \bq^{\prime}\}\rightarrow 0$ in Eq.(\ref{eq:snake}):
\ben
&&\lim_{\bq,\bq^{\prime}\rightarrow 0} {B}_3(\bq,\bq^{\prime},\bk,-\bk) \stackrel{\text{snake}}{=}
\rP(q)\rP(q^{\prime})\rP(k)\{\rF_2(\bq_1,\bk)\rF_2(-\bk,\bq_2) + \bq_1 \leftrightarrow \bq_2 \}.
\label{eq:contribution_snake}
\een
 
\n
{\bf Contributions From Star Diagrams:}
The following four terms represent the star contributions to trispectrum:
\ben
B_3(\bk_1,\bk_2,\bk_3,\bk_4) \stackrel{\text{star}}{=} \la \delta^{(3)}(\bk_1)\delta(\bk_2)\delta(\bk_3)\delta(\bk_4)\ra_c + \rm cyc.perm.
\label{eq:tri1_star}
\een
The expression for $\delta^{(3)}$ is expressed in terms of the kernel ${\rm F}_3$ defined in Eq.(\ref{eq:c21}):
\ben
&& \delta^{(3)}(\bk) = \delta_{3\rm D}(\bk-\bk_{abc})\int d^3\bk_a\,\delta({\bf k}_a) \int d^3\bk_b\,\delta({\bf k}_b) 
\int d^3\bk_c\,\delta({\bf k}_c)\; {\rm F}_3(\bk_a,\bk_b,\bk_c); \nn \\
&& \hspace{3cm} \bk_{abc}= \bk_a+\bk_b+\bk_c.
\een
We need to consider the following configuration in the squeezed limit:
\ben
&& \lim_{\bq_i\rightarrow 0} {B}_3(\bk_1-\bq_1,\bk_2-\bq_2,\bk_3-\bq_3, -\bq_4)\delta_{\rm 3D}(\bk_{123})\delta_{\rm 3D}(\bq_{1234});\nn \\
&& \hspace{3cm} \approx \lim_{\bq_4\rightarrow 0}{B}_3(\bk_1,\bk_2,\bk_3, -\bq_4)\delta_{\rm 3D}(\bk_{123}).
\een
The momentum-conserving Dirac's $\delta_{\rm 3D}$ function in the Fourier domain $\delta_{\rm 3D}(\bk_{123})$
reduced to $\delta_{\rm 3D}(\bk_{123})$ in the squeezed limit $\bq_4\rightarrow 0$. 
Thus effectively reducing the trispectrum to a bispectrum.
The terms that contribute are:
\ben
{B}_3(\bk_1,\bk_2,\bk_3,-\bq_4) \stackrel{\text{star}}{=}
{P}(q_4)\left [{\rm F}_3(\bk_1,\bk_2,-\bq_4){P}(k_1){P}(k_2) + {\rm cyc.perm.}\right ].
\een
Of the four terms listed in Eq.(\ref{eq:tri1_star}) only three survive as 
the contribution from the term ${\rm F}_3(\bk_1,\bk_2,\bk_3)$ vanishes due to the presence of
the factor $\delta_{\rm 3D}(\bk_{123})$. 
In the limit $\{\bq, \bq^{\prime}\}\rightarrow 0$ in Eq.(\ref{eq:tri1_star}):
\ben
&&\lim_{\bq,\bq^{\prime}\rightarrow 0} {B}_3(\bq,\bq^{\prime},\bk,-\bk) \stackrel{\text{star}}{=}
\rP(q)\rP(q^{\prime})\rP(k)
\left [\rF_3(\bq,\bq^{\prime},\bk)+\rF_3(\bq,\bq^{\prime},-\bk)  \right ].
\label{eq:contribution_star}
\een
{\bf Total Contribution:}
Combining contributions from both {\em star} and {\em snake} topologies we arrive at the following expression:
\ben
&& {B}_3(\bq,\bk_1,\bk_2,\bk_3) =
{P}(q_4)
\left [ \rF_2(\bk_1,\bk_2) P(k_1)P(k_2)+ {\rm cyc.perm.}\right ];\\
&& \rF_2(\bk_1,\bk_2) \equiv  \rF_3(\bk_1,\bk_2,\bq)+ {\rm F}_2(\bk_1,\bk_2) \left [ \rF_2(-\bq, \bk_1) + \bk_1 \rightarrow \bk_2 \right ].
\een
For the hierarchical anasatz $\la \rF_2 \ra = \nu_2$ and $\la \rF_3\ra = \nu_3$ and we get:
\ben
{B}_3(\bq,\bk_1,\bk_2,\bk_3) =
{P}(q) (\nu_3+2\nu_2^2) \left [ \rP(k_1)\rP(k_2)+\rP(k_2)\rP(k_3)+\rP(k_1)\rP(k_3)\right ].
\een
It thus takes an effective configuration of a bispectrum but with an amplitude determined by
coefficients that determine the trispectrum.

Combining expressions from Eq.(\ref{eq:contribution_snake}) 
and Eq.(\ref{eq:contribution_star}) we get in the limit $\{\bq, \bq^{\prime}\}\rightarrow 0$
\ben
&& \lim_{\bq,\bq^{\prime}\rightarrow 0} {B}_3(\bq,\bq^{\prime},\bk,-\bk) =
P(q)P(q^{\prime})P(k) \Big [\rF_3(\bq,\bq^{\prime},\bk)+\rF_3(\bq,\bq^{\prime},-\bk) \nn \\
&& \hspace{2cm} + \{\rF_2(\bq_1,\bk)\rF_2(-\bk,\bq_2) + \bq_1 \leftrightarrow \bq_2 \}\Big ].
\label{eq:total}
\een

\begin{figure}
\vspace{1.25cm}
\begin{center}
{\epsfxsize=10. cm \epsfysize=5. cm {\epsfbox[32 411 380 585]{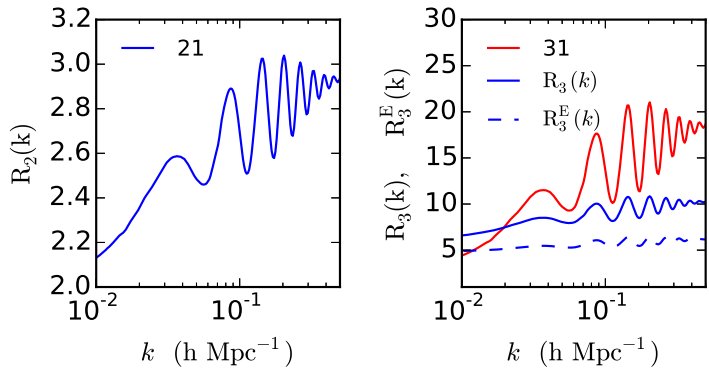}}}
\caption{The 3D normalised cumulant correlators [defined in Eq.(\ref{eq:c21})-Eq.(\ref{eq:c31})]
are compared with the coefficients $R_2$ and $R_3$ in Eq.(\ref{eq:R3})-Eq.(\ref{eq:R3E}) respectively.
The left panel shows $R_2$ and the right panel depicts $R_3$ and $R^{E}_3$ respectively
along with ${\rm C}_{31}$. Tree level perturbation theory is used in modelling of this
quantities. The quantities $R_2$ and $C_{21}$ are identical. However at third order
the coefficients $R_3$(or its Eulerian counterpart $R_3^{E}$) and ${\rm C}_{31}$ are different.}
\label{fig:cumu4}
\end{center}
\end{figure} 
\subsection{Squeezed-limit Trispectrum}
The integrated trispectrum (IT) $R_2(k)$ was derived in Ref.\citep{komatsu2} [see Eq.(A.19)]:
\ben
R_3(k)\stackrel{\text{tree}}{=} {8420 \over 1323} - {100 \over 63}
{d\ln P(k) \over d \ln k} + 
{1 \over 9}{k^2 \over P_{}(k)}{d^2 {}(k) \over dk^2}.
\een
To arrive at this result we have used the following expressions $\la \mu_1^2\ra=\la\mu_2^2 \ra ={1/3}$
and $\la\mu^2_{12}\ra=1/3$, $\la \mu_1\mu_2\mu_{12}\ra={1/9}$.
We will next use the following expression:
\ben
{k^2 \over P_{}(k)}{d^2 P_{}(k)\over dk^2} = \left [{d^2\ln P_{}(k) \over d (\ln k)^2} -{d\ln P_{}(k)\over d\ln k} + \left ( d\ln P_{}(k)\over d\ln k\right)^2\right ].
\een
To convert to Eulerian frame we use the following transformation in Eq.(4.1) of Ref.\citep{komatsu2} :
\ben
R_3^{\rm E}(k) = R_3(k) - 2\,f_2\,R_2(k); \quad f_2={17 \over 21}.
\een
\ben
R_3^{\rm E}(k) = && {8420 \over 1323} - {107 \over 63} \left [ {d\ln k^3 P_{}(k)\over d\ln k} -3 \right ]+
{1 \over 9}\left [{d\ln k^3 P_{}(k)\over d\ln k} -3 \right ]^2 + {1 \over 9}{d^2 \ln k^3 P_{}(k)\over d(\ln k)^2}\nn \\
&& -2 {17 \over 21}\left ( {68 \over 21} - {1\over 3}{d\ln k^3 P_{}(k) \over d \ln k} \right ).
\een
For a power law power spectra we have $\rP(k)\propto k^n$ and we have ${d\ln k^3 P(k)\over d\ln k} = (n+3)$
and the term involving the second derivative vanishes.

For $n=-3$ we have for $R_3$ and  $R^E_3$ we have:
\ben
\label{eq:r_3}
&& R_3^{}  \stackrel{\text{n=-3}}{=}{8420 \over 1323} + {107 \over 21} + {1}
= {16484\over 1323};\\
&& R_3^{\rm E} \stackrel{\text{n=-3}}{=} R_3^{}-2 \cdot{17 \over 21}\cdot {68 \over 21} 
{=} {1364\over 189}.
\een
Using HA we recover:
\ben
\lim_{\bq,\bq^{\prime}\rightarrow 0}  B_3[\bq,\bq^{\prime},\bk,-\bk] \stackrel{\text{\rm HA}}{=}
(4\nu_2^2+2\nu_3) \rP(q)\rP(q^{\prime})P(k).
\label{eq:my_sq}
\een
This is consistent with Eq.(\ref{eq:c31}) that defines the cumulant correlator $C_{31}$.

However, using PT kernels the results in Ref.\citep{komatsu2} are equivalent to
(for $n=-3$ in 3D):
\ben
\lim_{\bq,\bq^{\prime}\rightarrow 0}  B_3[\bq,\bq^{\prime},\bk,-\bk] \stackrel{\text{\rm PT}}{=}
(2\nu_2^2+2\nu_3) \rP(q)\rP(q^{\prime})P(k). 
\label{eq:their_sq}
\een
We get Eq.(\ref{eq:r_3}) if we use the squeezed limit in Eq.(\ref{eq:their_sq}).

In case of a locally power-law spectrum with arbitrary index $n$ we have:
\ben
R_3(k) \stackrel{\text{tree}}{=} {16484\over 1323}-{3129\over 1323}(n+3)+ {147 \over 1323}(n+3)^2.
\label{eq:R3}
\een
The Eulerican counterpart takes the following the expression:
\ben
R^{\rm E}_3(k) \stackrel{\text{tree}}{=}  {1364\over 189} -{345 \over 189}(n+3)+ {1 \over 9}(n+3)^2.
\label{eq:R3E}
\een
These expressions are plotted in Figure-\ref{fig:cumu4} along with their kurt-spectra counterpart defined in Eq.(\ref{eq:c31}).

\begin{thebibliography}{}
\bibitem{EW}
SDSS-III: Massive Spectroscopic Surveys of the Distant Universe, the Milky Way Galaxy, and Extra-Solar Planetary Systems
Eisenstein, D. J., Weinberg, D. H., Agol, E., et al. 2011, AJ, 142, 72,
[\href{http://lanl.arxiv.org/abs/1101.1529}{\tt arXiv:1101.1529}]
\bibitem{DJA}
The WiggleZ Dark Energy Survey: Survey Design and First Data Release, 
Drinkwater, M. J., Jurek, R. J., Blake, C., et al. 2010, MNRAS, 401, 14,
[\href{http://lanl.arxiv.org/abs/0911.4246}{\tt arXiv:0911.4246}]
\bibitem{DES}
The Dark Energy Survey,
The Dark Energy Survey Collaboration 2005, arXiv:astro-ph/0510346,
[\href{http://arxiv.org/abs/astro-ph/0510346}{{\tt astro-ph/0510346}}].
\bibitem{LAA}
Euclid Definition Study Report
Laureijs, R., Amiaux, J., Arduini, S., et al. 2011, arXiv:1110.3193,
[\href{http://arxiv.org/abs/arxiv:1110.3193}{{\tt arXiv:1110.3193}}].
\bibitem{skewspec}
A New Approach to Probing Primordial Non-Gaussianity,
Munshi D., Heavens A.,
2010, MNRAS, 401, 2406,
[\href{http://arxiv.org/abs/0904.4478}{{\tt arXiv/0904.4478}}].
\bibitem{kurtspec}
New Optimised Estimators for the Primordial Trispectrum,
Munshi D., Heavens A., Cooray A., Smidt J., Coles P., Serra P.,
2011, MNRAS, 412,1993,
[\href{http://arxiv.org/abs/0910.3693}{{\tt arxiv/0910.3693}}].
\bibitem{MunshiWaerbeke}
From Weak Lensing to non-Gaussianity via Minkowski Functionals
Munshi D., van Waerbeke L., Smidt J., Coles P.,
2012, MNRAS, 419, 536
\bibitem{review}
Large scale structure of the
universe and cosmological perturbation theory,
Bernardeau F., Colombi S., Gaztanaga E., Scoccimarro R.,  2002, Phys.Rept. 367, 1,
[\href{http://lanl.arxiv.org/abs/astro-ph/0112551}{{\tt astro-ph/0112551}}].
\bibitem{EFT}
The Effective Field Theory of Cosmological Large Scale Structures
Joseph J., Carrasco M., Hertzberg M.P., Senatore L.
JHEP, Volume 2012, Number 9 (2012), 82,
[\href{http://arxiv.org/abs/1206.2926}{{\tt arXiv:1206.2926}}]
\bibitem{halo}
Halo Models of Large Scale Structure
Cooray A., Sheth R. 2002, Phys.Rept.372, 1,
[\href{http://arxiv.org/abs/astro-ph/0206508}{{\tt astro-ph/0206508}}]
\bibitem{komatsu}
Position-dependent power spectrum of the large-scale structure: a novel method to measure the squeezed-limit bispectrum,
Chiang C-T, Wagner C.,Schmidt F., Komatsu E.,
[\href{http://lanl.arxiv.org/abs/1403.3411}{{\tt arXiv/1403.3411}}].
\bibitem{francis}
The large-scale Gravitational Bias from the Quasilinear Regime,
Bernardeau F., 1996, A\&A 312, 11 
[\href{http://lanl.arxiv.org/abs/astro-ph/9602072}{{\tt arXiv/9602072}}].
\bibitem{gen}	
Generalised Cumulant Correlators and Hierarchical Clustering,
Munshi D.; Melott A. L., Coles P., 2000, MNRAS, 311, 149
[\href{http://lanl.arxiv.org/abs/astro-ph/9812271}{{\tt arXiv/9812271}}].
\bibitem{inf1}
The Separate Universe Approach to Soft Limits,
Kenton Z., Mulryne D.J.,
[\href{http://arxiv.org/abs/1605.03435}{{\tt arXiv/1605.03435}}]
\bibitem{valag1}
Kinematic consistency relations of large-scale structures,
Valageas P., 2014, PRD, 89, 083534
[\href{http://arxiv.org/abs/1311.1236}{{\tt arXiv/1311.1236}}]
\bibitem{valag2}
Angular averaged consistency relations of large-scale structures, 
Valageas P., 2014, PRD, 89, 123522, 
[\href{http://lanl.arxiv.org/abs/1311.4286}{{\tt arXiv/1311.4286}}]
\bibitem{nishi1}
Testing the equal-time angular-averaged consistency relation of the gravitational dynamics in N-body simulations
Nishimichi T., Valageas P., 2014, PRD, 90, 023546, 
[\href{http://arxiv.org/abs/1402.3293}{{\tt arXiv/1402.3293}}]
\bibitem{hypPT}
Hyperextended Cosmological Perturbation Theory: Predicting Non-linear Clustering Amplitudes,
Scoccimarro R., Frieman J.A., 1999, ApJ, 520, 35
[\href{http://arxiv.org/abs/astro-ph/9811184}{{\tt astro-ph/9811184}}].
\bibitem{SzSz}
Cumulant Correlators from the APM, 
Szapudi I., Szalay A.S., 
1999,  Astrophys.J., 515, L43,
[\href{http://arxiv.org/abs/astro-ph/9702015}{{\tt astro-ph/9702015}}].
\bibitem{MC00}
Weak lensing from strong clustering,	
Munshi, D; Coles, P, 2000, MNRAS, 313, 148,
[\href{http://lanl.arxiv.org/abs/astro-ph/9911008}{{\tt astro-ph/9911008}}].
\bibitem{DP77}
On the integration of the BBGKY equations for the development of strongly nonlinear clustering in an expanding universe,
Davis M., Peebles, P.J.E. 1977, ApJS, 34, 425,
[\href{http://adsabs.harvard.edu/cgi-bin/bib_query?1977ApJS...34..425D}{{\tt 1977 ApJS 34 425D}}]
\bibitem{GP77}
Statistical Analysis Of Catalogs Of Extragalactic Objects. VII -Two- And -Three- Point 
Correlation Functions For The High-Resolution Shane-Wirtanen Catalog Of Galaxies,
Groth E., Peebles, P.J.E., 1977, ApJ, 217, 385,
[\href{http://adsabs.harvard.edu//abs/1977ApJ...217..385G}{{\tt 1977 ApJ 217 385G}}]
\bibitem{Fry84}
The Galaxy correlation hierarchy in perturbation theory
Fry J.N., 1984b, ApJ, 279, 499,
[\href{http://adsabs.harvard.edu/abs/1984ApJ...279..499F}{{\tt 1984 ApJ 279 499F}}]
\bibitem{Ber92}	
The gravity-induced quasi-Gaussian correlation hierarchy
Bernardeau, F. 1992, ApJ, 192, 1,
[\href{http://adsabs.harvard.edu/abs/1992ApJ...392....1B}{{\tt 1992 ApJ 392 1B}}]
\bibitem{Ber94}
The Effects of Smoothing on the Statistical Properties of the Large-Scale Cosmic Fileds 
Bernardeau, F. 1994, A\&A, 291, 697,
[\href{http://arxiv.org/abs/astro-ph/9403020}{{\tt astro-ph/9403020}}].
\bibitem{FryPeebles78}
Statistical analysis of catalogs of extragalactic objects. IX - The four-point galaxy correlation function
Fry J.N., Peebles P.J.E., 1978, ApJ, 221, 19,
[\href{http://adsabs.harvard.edu/abs/1978ApJ...221...19F}{\tt 1978 ApJ 221 19F}]
\bibitem{Peebles80}
Peebles, P.J.E. 1980, The Large Scale Structure of the Universe, 
Princeton University Press, Princeton, N.J., USA
\bibitem{BaS99}
Scale-invariant matter distribution in the universe. I - Counts in cells,
Balian R., Schaeffer R., 1989, A\&A, 220, 1
\bibitem{newHam88b}
Hamilton, A.J.S., 1988b, ApJ, 332, 67
\bibitem{HKLM91}
Reconstructing the primordial spectrum of fluctuations of the universe from the observed nonlinear clustering of galaxies.
Hamilton, A.J.S., Kumar, P., Lu, E., Mattews, A. 1991, ApJ, 274, 1; Erratum: 1995, ApJ, 442L, 73H,
[\href{http://adsabs.harvard.edu/abs/1991ApJ...374L...1H}{1991 ApJ 374L 1H}]
\bibitem{BeS99}
Halo correlations in nonlinear cosmic density fields.
Bernardeau F., Schaeffer R., 
1999, A\&A, 349, 697B,
[\href{http://arxiv.org/abs/9903087}{{\tt astro-ph/990387}}].
\bibitem{BeS92}
Galaxy correlations, matter correlations and biasing,
Bernardeau F., Schaeffer R., 
1992, A\&A, 255, 1
\bibitem{bias}
Bias and Hierarchical Clustering,
Coles P., Melott A.L., Munshi D.,  1999, ApJ, 521L, 5C,
[\href{http://lanl.arxiv.org/abs/astro-ph/9904253}{{\tt astro-ph/9904253}}]
\bibitem{snst1}
From Snakes to Stars, the Statistics of Collapsed Objects - I. Lower-order Clustering Properties,
Munshi D., Coles P., Melott A.L.
1999, MNRAS, 307, 387,
[\href{http://lanl.arxiv.org/abs/astro-ph/9812337}{{\tt astro-ph/9812337}}]
\bibitem{snst2}
From Snakes to Stars, the Statistics of Collapsed Objects - II. Lower-order Clustering Properties,
Munshi D., Coles P., Melott A.L.,
1999, MNRAS, 310, 892,
[\href{http://lanl.arxiv.org/abs/astro-ph/9902215}{{\tt astro-ph/9902215}}]
\bibitem{scaling}
Scaling in Gravitational Clustering, 2D and 3D Dynamics,
Munshi D., Bernardeau F., Melott A.L., Schaeffer R., 1999,
MNRAS, 303, 433,
[\href{http://lanl.arxiv.org/abs/astro-ph/9707009}{{\tt astro-ph/9707009}}]
\bibitem{Peebles}
The Effect of a Lumpy Matter Distribution on the Growth of Irregularities in
an Expanding Universe, P. J. E. Peebles, 1974, A\&A, 32, 391.
\bibitem{theta_skew}
Omega from the skewness of the cosmic velocity divergence,
Bernardeau F., Juszkiewicz R., Dekel A., Bouchet F.R., 
1995, MNRAS, 274, 20,
[\href{http://lanl.arxiv.org/abs/astro-ph/9404053}{{\tt astro-ph/9404052}}].
\bibitem{Scoccimarro}
Scoccimarro R., Couchman, H.M. 
2011, MNRAS, 325, 1312, 
[\href{http://lanl.arxiv.org/abs/0009427}{{\tt astro-ph/009427}}]
\bibitem{GilMarin}
An improved fitting formula for the dark matter bispectrum
Gil-Marín H., Wagner C., Fragkoudi F., Jimenez R., Verde L.,
2012, JCAP, 02, 047
[\href{http://lanl.arxiv.org/abs/1111.4477}{{\tt astro-ph/111.4477}}]
\bibitem{Baldauf}
Galaxy Bias and non-Linear Structure Formation in General Relativity, 
T. Baldauf, U. Seljak, L. Senatore, and M. Zaldarriaga, 2011, JCAP, 1110, 031
[\href{http://lanl.arxiv.org/abs/astro-ph/1106.5507}{{\tt arXiv/1106.5507}}].
\bibitem{Creminelli}
Single-Field Consistency Relations of Large Scale Structure
Creminelli P, Noreña J., Simonović M., Vernizzi F.
2013, JCAP, 12, 025C
[\href{http://lanl.arxiv.org/abs/1309.3557}{{\tt arxiv/1309.3557}}]
\bibitem{WayneHu}
Super-Sample Covariance in Simulations
Li Y., Hu W., Takada M., 2014, PRD, 89, 083519 
[\href{http://lanl.arxiv.org/abs/1401.0385}{{\tt arxiv/1401.0385}}]
\bibitem{Pajer}
The Observed squeezed limit of cosmological three-point functions,
E. Pajer, F. Schmidt, and M. Zaldarriaga,  2013, PRD, 88, 083502, 
[\href{http://lanl.arxiv.org/abs/astro-ph/1305.0824}{{\tt arXiv/1305.0824}}].
\bibitem{MVS}
Munshi, D., Sahni, V., Starobinsky, A. A.,
Nonlinear approximations to gravitational instability: A comparison in the quasi-linear regime,
1994, ApJ, 436, 517M [\href{http://arxiv.org/abs/9402065}{{\tt astro-ph/9402065}}].
\bibitem{TakadaJain}
Cosmological parameters from lensing power spectrum and bispectrum tomography
Takada M, Jain B., 2004, MNRAS, 348, 897
[\href{http://lanl.arxiv.org/abs/astro-ph/0310125}{{\tt arxiv/0310125}}]
\bibitem{2Dcumu}
Weak Lensing Statistics as a Probe of Omega and Power Spectrum,
F. Bernardeau, L. van Waerbeke, Y. Mellier, 1997, A\&A. 322, 1, 
[\href{http://arxiv.org/abs/9609122}{{\tt astro-ph/9609122}}].
\bibitem{Bernard_angular}
The angular correlation hierarchy in the quasilinear regime,
Bernardeau F., 1995, A\&A, 301, 309
[\href{http://lanl.arxiv.org/abs/astro-ph/9502089}{{\tt arXiv/9502089}}].
\bibitem{PrattenMunshi}
Non-Gaussianity in large-scale structure and Minkowski functionals
Pratten G., Munshi D., 2012, MNRAS, 423, 3209,
[\href{http://lanl.arxiv.org/abs/1108.1985}{{\tt arxiv/1108.1985}}]
\bibitem{komatsu2}
The angle-averaged squeezed limit of nonlinear matter N-point functions
Wagner C, Schmidt F., Chiang C.-T.,  Komatsu E.,
2015, JCAP, 08, 042,
[\href{http://arxiv.org/abs/1503.03487}{{\tt arxiv/1503.03487}}]
\bibitem{galaxy_red}
Galaxy clustering in 3D and modified gravity theories,
Munshi D., Pratten G., Valageas P., Coles P., Brax Ph.,
2016, MNRAS, 456, 1627
[\href{http://lanl.arxiv.org/abs/1508.00583}{{\tt arxiv/1508.00583}}]
\bibitem{Planck}
Planck 2015 results. XVII. Constraints on primordial non-Gaussianity 
Planck Collaboration
[\href{http://lanl.arxiv.org/abs/1502.01592}{{\tt arxiv/1502.01592}}]
\bibitem{3Dlens1}
3D Weak Lensing: Modified Theories of Gravity
Pratten G., Munshi D., Valageas P., Brax Ph., 
2016, PRD, 93, 103524,
[\href{http://lanl.arxiv.org/abs/1602.06711}{{\tt arxiv/1602.06711}}]
\bibitem{3Dlens2}
Higher order statistics for three-dimensional shear and flexion	
Munshi D., Kitching T., Heavens A., Coles P.,
2011, MNRAS, 416, 1629,
[\href{http://lanl.arxiv.org/abs/1012.3658}{{\tt arxiv/1012.3658}}]
\bibitem{sq_cmb}
CMB lensing and primordial squeezed non-Gaussianity,
Pearson R., Lewis A., Regan D., 2012, JCAP, 03, 011,
[\href{http://arxiv.org/abs/1201.1010}{{\tt arXiv:1201.1010}}]
\bibitem{sq_cmb1}
The full squeezed CMB bispectrum from inflation,
Lewis A.,
2012, JCAP, 06, 023, [\href{http://arxiv.org/abs/1204.5018}{{\tt arXiv:1204.5018}}]
\bibitem{CMB_lensing}
Lensing-induced morphology changes in CMB temperature maps in modified gravity theories,
Munshi D., Hu B., Matsubara T., Coles P., Heavens A., 
2016, JCAP, 04, 056
[\href{http://lanl.arxiv.org/abs/1602.00965}{{\tt arxiv/1602.00965}}]
\bibitem{kSZ}
Extracting the late-time kinetic Sunyaev-Zel'dovich effect,
D. Munshi, I. T. Iliev, K. L. Dixon, P. Coles
[\href{http://lanl.arxiv.org/abs/1511.034495}{{\tt arxiv/1511.034495}}]
\bibitem{tSZ1}
Cross-correlating Sunyaev-Zel'dovich and weak lensing maps,
Munshi D., Joudaki S., Coles P., Smidt J., Kay S. T.,
2014, MNRAS, 442, 69 
[\href{http://lanl.arxiv.org/abs/1111.5010}{{\tt arxiv/1111.5010}}]
\bibitem{Pandolfi}
Reionization and CMB non-Gaussianity
Munshi D., Corasaniti P. S., Coles P., Heavens A., Pandolfi S. 
[\href{http://lanl.arxiv.org/abs/1403.1531}{{\tt arXiv/1403.1531}}]
\bibitem{tSZ2}
Statistical Properties of Thermal Sunyaev-Zel'dovich Maps,
Munshi D., Joudaki S., Smidt J., Coles P., Kay S. T.,
2013, MNRAS, 429, 1564 
[\href{http://lanl.arxiv.org/abs/1106.0706}{{\tt arxiv/1106.0706}}]
\bibitem{valag_consistency}
Redshift-space equal-time angular-averaged consistency 
relations of the gravitational dynamics,
Nishimichi T., Valageas P.,
2015, PhRvD,92, 123510,
[\href{http://lanl.arxiv.org/abs/1503.06036}{{\tt arxiv/1503.06036}}]

















\end{thebibliography}
\end{document}